\documentclass[%
 reprint,
 superscriptaddress, 
 amsmath,amssymb,
 aps,
 pra,
 floatfix,
]{revtex4-2}

\usepackage{graphicx}
\usepackage{dcolumn}
\usepackage{bm}
\usepackage{dsfont}
\usepackage{xcolor}
\usepackage{enumitem}
\usepackage{multirow}
\usepackage{hyperref}
\usepackage{array}
\usepackage{float}
\usepackage{csquotes}
\usepackage[T1]{fontenc}
\MakeOuterQuote{"}

\usepackage[textsize=tiny]{todonotes}

\usepackage{braket}
\usepackage{listings}
\usepackage{xcolor}
\usepackage{appendix}

\newcommand{\outprod}[1]{\ket{#1}\!\!\bra{#1}}

\newcommand{\inprod}[2]{\Braket{#1|#2}}
\newcommand{\mmnt}[1]{\langle{#1}\rangle}
\DeclareMathOperator{\Tr}{Tr}

\def\GR{\textcolor{black}}

\definecolor{customcolorblue}{HTML}{4573ae}

\begin{document}

\title{Full-stack Physics-level model of cascaded entanglement links}

\author{J. Gabriel Richardson}
\email[]{rgb@umd.edu}
\affiliation{Department of Electrical and Computer Engineering, University of Maryland, College Park, MD 20742 }
\affiliation{NSF-ERC Center for Quantum Networks, The University of Arizona, Tucson, AZ 85721}

\author{Prajit Dhara}
\altaffiliation{Now at: RTX BBN Technologies, Cambridge, MA}
\affiliation{Department of Electrical and Computer Engineering, University of Maryland, College Park, MD 20742 }
\affiliation{NSF-ERC Center for Quantum Networks, The University of Arizona, Tucson, AZ 85721}

\author{Abhishek Bhatt}
\affiliation{College of Information and Computer Science, University of Massachusetts Amherst}
\affiliation{NSF-ERC Center for Quantum Networks, The University of Arizona, Tucson, AZ 85721}

\author{Saikat Guha}
\affiliation{Department of Electrical and Computer Engineering, University of Maryland, College Park, MD 20742 }
\affiliation{NSF-ERC Center for Quantum Networks, The University of Arizona, Tucson, AZ 85721}

\author{Stefan Krastanov}
\email{skrastanov@umass.edu}
\affiliation{College of Information and Computer Science, University of Massachusetts Amherst}
\affiliation{Department of Physics, University of Massachusetts Amherst}
\affiliation{NSF-ERC Center for Quantum Networks, The University of Arizona, Tucson, AZ 85721}

\date{\today}

\begin{abstract}

While the last few decades have seen a proliferation of experimental demonstrations of entanglement sources, practicality of deployment has been a secondary concern. Recently, the ZALM source~\cite{chen_zero-added-loss_2023} was introduced, as a well-engineered functional device, easily integrated within a complete networking system. It addresses numerous concerns which make typical academic demonstrations less practical: reliable heralding signals, multiplexing across multiple dimensions, and efficient use of input power. We present a stack of tools for modeling mode-by-mode a ZALM source under realistic conditions, in isolation or as a part of a complete network testbed. Our modeling formalism builds upon a hybrid Gaussian and non-Gaussian representation, providing a flexible tradeoff between performance and accuracy, while also greatly simplifying the exact calculation of otherwise expensive scalar figures of merit. This toolkit, implemented in the Python package called "genqo", is integrated within the QuantumSavory full-stack simulator and the QuantumSymbolics computer algebra system. We use this software stack to demonstrate a number of complete networking protocols built upon the ZALM source.

\end{abstract}

\maketitle

\section{Introduction}

The gasoline to the engine of quantum networks is entanglement, particularly photonic entanglement that can be transmitted through existing optical fiber infrastructure or via free-space satellite links. This entanglement, generated by a source, may be used in its photonic format, but for many quantum computing and sensing applications it will be necessary to transduce it to some form of quantum memory, \GR{which is} a qubit capable of universal quantum logic with enough coherence time for the specific application. The applications, such as distributed quantum computing~\cite{fitzsimons_private_2017, wehner_quantum_2018}, entanglement enhanced long-baseline interferometry~\cite{gottesman_longer-baseline_2012, khabiboulline_quantum-assisted_2019,khabiboulline2019optical}, or entangled sensor networks ~\cite{van_milligen_utilizing_2024, zhang_distributed_2021} are hungry for entanglement as distillation protocols generate entangled states in their most useful and pure form by combining received noisy Bell pairs. As such, it is necessary for entanglement sources to produce this resource at high rates and with high quality or fidelity. 

Previous proposals for entanglement sources have a number of bottlenecks. Their fundamental mechanism for generating entanglement is via optical nonlinear processes, such as spontaneous parametric down conversion (SPDC). Colloquially called "SPDC sources" ~\cite{kwiat_new_1995, kwiat_ultrabright_1999}, these sources use engineered nonlinear optical crystals which are pumped by pulsed laser systems. The result is that with some probability, a single pair of entangled photons will be produced. \GR{Quantum memories are required to herald when an entangled photon pair is produced, and} thus ultimate entanglement generation rates of these sources are limited, even suffering from a large memory resource overhead when performing frequency multiplexing. 

\begin{figure}[h]
    \centering
    \includegraphics[width=\linewidth]{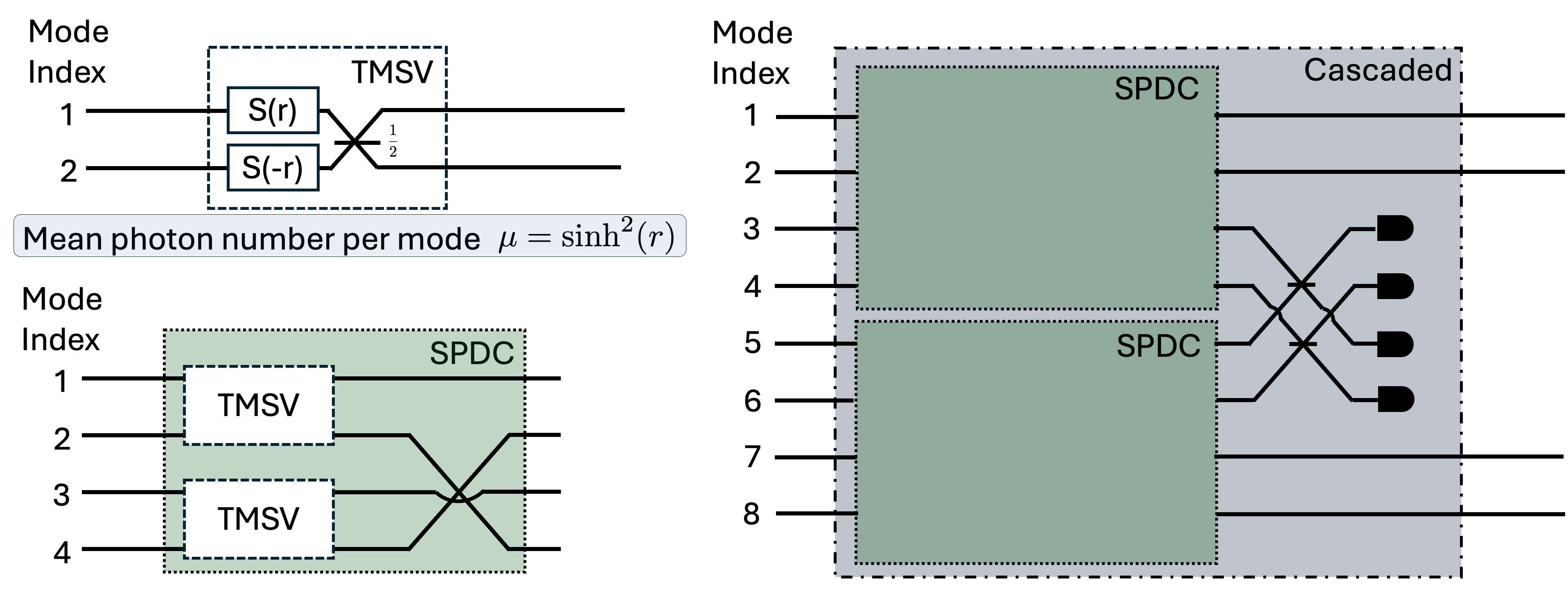}
    \caption{A diagram showing the construction of \GR{the cascaded/ZALM source from the SPDC source, which is constructed of two mode squeezed vacuum (TMSV) states, which are composed of single-mode squeezing operations follow by a beam splitter. The squeezing parameter, $r$, is related to the mean photon number, $\mu = \sinh^2(r)$, which is the fundamental operational parameter. Typically, $\mu \approx 0.1$, and thus $r = \text{arcsinh}(\sqrt{0.1}) \approx 0.31$.} The heralding mechanism of the cascaded source via an entanglement swap between half of the modes from two SPDC sources is essential for ZALM operation. This enables frequency multiplexing without introducing a large quantum memory overhead, as would be required for frequency multiplexing unheralded SPDC source. Cascading also enables source operation in a higher mean photon number regime as compared to the SPDC source, since the heralding mechanism essentially filters undesirable higher-photon-order terms. \GR{Leading theoretical ZALM proposals anticipate high-fidelity entanglement rates on the order of tens of megahertz.} \cite{shapiro_high-fidelity_2025}}
    \label{fig:diagram}
\end{figure}

Thus, a different concept emerges: the cascaded-heralded entanglement source~\cite{dhara_heralded_2022}. This source is realized by utilizing two SPDC sources and performing a heralding Bell state measurement between two modes from each source. In this configuration, the cascaded source overcomes the challenge of the receiver not knowing when an entangled pair is produced. But, this heralding will reduce the rate of entanglement produced. An extension of the cascaded source, called the zero-added loss multiplexing (ZALM) source~\cite{chen_zero-added-loss_2023}, proposes using frequency multiplexing to produce entangled photon pairs quasi-deterministically over some frequency band. Together with frequency mode conversion, this source overcomes the bottlenecks of other entanglement source proposals, drafting a blue print for potentially one of the most important operating technologies of tomorrow's quantum networks. Figure \ref{fig:diagram} shows from the perspective of operations on the optical modes how the cascaded (a.k.a.\ ZALM) source is constructed from two SPDC sources.

The development and engineering of entangled photonic qubit sources based on ZALM is an active area of research with practical engineering challenges that need to be solved for practical implementations over realistic network links. Initial analyses of ZALM sources were limited in that they truncated the quantum state using a low mean photon number approximation~\cite{shapiro_entanglement_2024, chen_zero-added-loss_2023}, thus limiting analysis of source performance to the regime where the pump laser power is low.  A technique for overcoming this approximation by using a hybrid Gaussian/non-Gaussian source model was proposed in an article about cascaded sources ~\cite{dhara_heralded_2022}; however, this technique was not pursued. A similar approach was recently used to analyze a ZALM source that utilizes phase-matched spectral islands~\cite{shapiro_high-fidelity_2025}, providing further evidence to support the near-term realization of a ZALM source with entanglement rates necessary for high-demand applications.  

In this paper, we present a detailed mathematical approach to modeling entanglement sources based on photonic nonlinearities. We analyze the SPDC source and the cascaded/ZALM source using this hybrid Gaussian/non-Gaussian approach, thereby removing problematic inaccuracies that limited the parameter regimes studied in previous work~\cite{shapiro_entanglement_2024}. We include formalism for calculating both the photon-photon state of the source, in the presence of non-idealities such as loss and detector excess noise, as well as the spin-spin state, in the presence of the same non-idealities, following loading into idealized Duan-Kimble~\cite{duan_scalable_2004} style quantum memories. We also present an open-source toolbox, "genqo", which implements this formalism in a Python modeling package. 

The more accurate models presented here, uncover parameter regimes of interest, providing an opportunity for much higher overall rate and fidelity of distributing entangled pairs. As shown in this paper, at higher pump intensities, we have access to higher success probabilities that previous approximate models did not predict.

Next-generation quantum networks will not exist without high rates of entanglement generation, and ZALM is a proposed leading solution to this challenge. Our tools presented herein can enable ZALM architects to properly understand the source and its components in the context of the system. 

This paper is structured as follows. In \textit{"Physical Principles"}, we give a schematic description of the entanglement source\GR{s} under consideration, reiterate its advantages, and present the experimental parameters that define the source and the critical figures of merit to evaluate. In \textit{"Mathematical Principles"}, we provide a review of the techniques necessary for the successful high-fidelity modeling of the\GR{se} entanglement source\GR{s}. References to software implementations of each of these concepts (within the genqo framework) are provided. Then, in the \textit{"Simulation Module"}, we present a few direct use cases of the genqo framework and demonstrate how to employ it for studying entanglement sources. The section \GR{contains} a discussion and examples of how the genqo framework can be used within the full-stack networking simulator QuantumSavory.jl or the computer algebra system QuantumSymbolics.jl.

\section{Physical Principles}

Spontaneous parametric down conversion (SPDC) is a nonlinear optical process that takes an input photon (colloquially called the pump) and produces two output photons (colloquially called the signal and idler) \cite{Boyd_2003}. The photons satisfy conservation of energy and momentum by the relationships \cite{Ou_2007}
\begin{align}
    \omega_P = \omega_S + \omega_I \\
    \mathbf{k}_p = \mathbf{k}_S + \mathbf{k}_I
\end{align}

We consider the action of a two-mode squeezing Hamiltonian $\hat{H}_{\rm TMS}$ generated by a spontaneous parametric down conversion (SPDC) process on the signal and idler fields (labeled by the subscripts $S$ and $I$)
\begin{widetext}
\begin{align}
    \hat{H}_{\rm TMS} (\xi) = \iint \, d\omega_S\, d \omega_I \left(\xi \Psi(\omega_S,\omega_I) \hat{a}^\dagger_{S}(\omega_S) \hat{a}^\dagger_I(\omega_I)  - \xi^* \Psi(\omega_S,\omega_I)^* \hat{a}_{S}(\omega_S) \hat{a}_{I}(\omega_I)\right)
\end{align}
\end{widetext}
where $\xi$ governs the strength of the two-mode interaction, and $\Psi(\omega_S,\omega_I) $ is the joint-spectral amplitude (JSA) function that governs the spectral correlations in the signal and idler beams. 

Starting from the Hamiltonian for a down conversion process in terms of classical fields followed by quantizing the electric field \cite{grice_spectral_1997}, the joint spectral amplitude (JSA) $\Psi(\omega_S, \omega_I)$ arises as a function of the properties of the pump field as well as the nonlinear crystal. It represents the probability amplitude of generating two photons; one at $\omega_S$ and one at $\omega_I$, such that
\begin{align}
    \Psi(\omega_S, \omega_I) = \alpha(\omega_S + \omega_I) \Phi(\omega_S, \omega_I)
\end{align}
where $\alpha(\omega_P)$ is called the spectral mode shape function, which the ZALM paper \cite{chen_zero-added-loss_2023} considers as Gaussian having the form $\alpha(\omega_P) = \exp(-(\omega_P - \omega_{P0})^2/2\sigma^2_P)$ where $\omega_{P0}$ is the pump center frequency and $\sigma_P$ is the pump linewidth; $\Phi(\omega_S,\omega_I)$ is called the Phase Matching Function, which depends on the physical characteristics of the nonlinear crystal. This description provides "knobs" for engineering the properties of the photons produced by the down conversion process, either by changing the characteristics of the pump or of the crystal. These "knobs" have been considered in previous studies, such as by Shapiro et al \cite{shapiro_entanglement_2024}.

\begin{figure}[h]
    \centering
    \includegraphics[width=\linewidth]{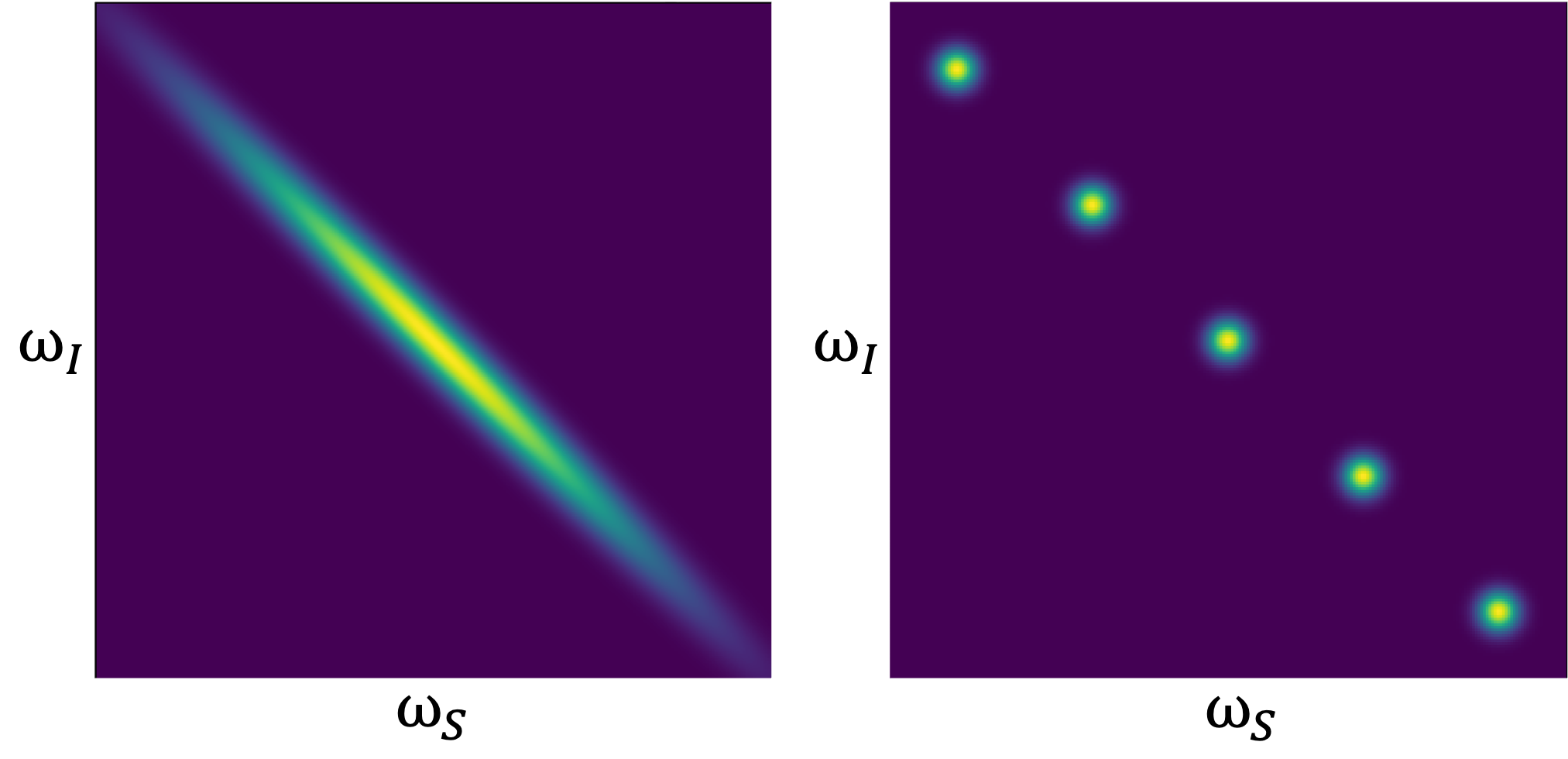}
    \caption{\GR{Two joint spectral amplitude (JSA) configurations for multiplexed ZALM sources. The JSA on the left is sometimes referred to as a bi-Gaussian JSA, while the one on the right is sometimes referred to as an island JSA.}}
    \label{fig:jsa_pic}
\end{figure}

In this paper, we present the modeling process for a single frequency mode. To model frequency multiplexing, we must consider the specific structure of the JSA. \GR{Figure \ref{fig:jsa_pic} shows two example JSA's, the bi-Gaussian and the island JSA. Because the island JSA is fully seperable \cite{shapiro_high-fidelity_2025} it can be completely modeled by \texttt{genqo} as the total mean photon number is distributed to each island evenly. On the other hand, the performance of the bi-Gaussian JSA will depend on the filtering mechanism used for separating the frequency modes, introducing effects such as modal cross talk, which can not be considering using \texttt{genqo} at present, but remains an open action for future research.} 

The SPDC process occurs in different types (0,I,II) depending on the group indices of the signal and idler photons. The frequency of the produced photons are either degenerate (the signal and idler have the same frequency) or non-degenerate (the signal and idler do not have the same frequency)~\cite{schneeloch_introduction_2019}. The quantum state produced by a non-degenerate \GR{SPDC process} is the two mode squeezed vacuum (TMSV) \GR{state}~\cite{lvovsky_squeezed_2016}, represented in the Fock basis is~\cite{schumaker_new_1985}

\begin{align}
    S(r,\theta) \ket{0} = (\cosh{r})^{-1} \sum_{n = 0}^{\infty} (-e^{2 i \phi} \tanh{r})^n \ket{n} \ket{n}
\end{align}
Where $r$ is a real number called the squeezing factor, and $\phi$ is a phase angle. The \textit{mean photon number per mode}, $\mu$ is related to $r$ by $\mu = \sinh^2(r)$ \cite{lvovsky_squeezed_2016}. Since $\cosh^2(r) - \sinh^2(r) = 1$, $\cosh(r) = \sqrt{1 + \sinh^2(r)} = \sqrt{1 + \mu}$. Also, $\tanh(r) = \frac{\sinh(r)}{\cosh(r)} = \sqrt{\frac{\mu}{1 + \mu}}$. So, the two mode squeezed vacuum state can be written as
\begin{align}
    \ket{\text{TMSV}} &= \frac{1}{\sqrt{1 + \mu}} \sum_{n = 0}^{\infty} \left(\frac{\mu}{1 + \mu}\right)^{n/2} \ket{n} \ket{n} \\
    &= \sum_{n = 0}^{\infty} \sqrt{\frac{\mu^{n}}{(1 + \mu)^{n+1}}}\ket{n} \ket{n}
\end{align}

An SPDC source is constructed from two SPDC processes, or mathematically two, two-mode squeezed vacuum states with an idler mode swap. The first source of polarization entangled photons pairs was created using two non-linear optical crystals, rotated $90 \deg$ with respect to each other \cite{kwiat_new_1995, kwiat_ultrabright_1999}. Today, a commonly used configuration of this source is the Sagnac configuration~\cite{kim_phase-stable_2006, wong_efficient_2006}. 

\subsubsection{The Heralding Process}

As mentioned previously, the cascaded (i.e., ZALM) source is constructed from two SPDC sources, where a Bell state measurement is performed on a pair of modes (one from each SPDC source) to herald the generation of dual-rail entangled photons. The dual-rail Bell state which we seek to produce from these sources has the form $\frac{1}{\sqrt{2}} (\ket{1,0;0,1} \pm \ket{0,1;1,0})$ where the bases of the state might be two spatial modes, each with a horizontal or vertical polarizaiton (as is the case in the Sagnac configuration). The state produced by the SPDC source will have a vacuum component, as well as multi-photon pair terms \cite{dhara_heralded_2022}. These multi-photon pair terms are generally not useful for the receivers, and are considered failure. Using the Bell state measurement, which in the dual-rail case would require click patterns where one horizontally polarized photon is measured and one vertically polarized photon is measured after beam splitter operations between the respective modes, we are able to essentially filter the multi-photon pair terms. The resulting state of the cascaded source is either the Bell state portion (as desired) or two photon terms $\ket{1,1;0,0}$ or $\ket{0,0;1,1}$ where both photons arrive at one receiver. This state is much more straightforward from a utility perspective, particularly in the presence of quantum memories. 

Figure \ref{fig:diagram} shows the process of constructing the source from the perspective of operations on the optical modes. We begin with four TMSV states. An idler mode swap occurs between pairs of TMSVs, followed by a Bell state measurement between two modes from each resulting SPDC source which is accomplished by interfering the corresponding mode from that of the other source, and single photon detection. 

\subsubsection{Using Quantum Memories}

Duan-Kimble style quantum memories~\cite{duan_scalable_2004} realize the operation of tranducing a dual-rail qubit into some atomic qubit basis state deterministically, such that
\begin{align}
    \alpha \ket{1,0} + \beta \ket{0,1} \rightarrow \alpha \ket{\mathbf{0}} + \beta \ket{\mathbf{1}}
\end{align}
Where $\ket{\mathbf{0/1}}$ are the basis states of the atomic qubit. This is realized by: (1) a photon reflection off of the atom-cavity system, such that a phase is aquired if the photon is in one optical mode, but not in the other. In quantum computing terms, this is a CZ operation; (2) Interference of the two optical modes via a beam splitter; (3) single-photon detection in one of the optical modes. Using quantum memories, we can in theory herald the reception of a dual-rail Bell state from the entanglement source, and filter out the situations where two photons arrived at one receiver and not at the other. 

\subsubsection{Definition of Performance Metrics}

We consider two primary performance metrics: (1) $P_{gen}$, \textit{the probability of generation}, which for the cascaded source case is the probability that a successful Bell state measurement occurs. For the SPDC source, the probability of generation might be calculated in the presence of memories, and would be the probability that a Bell state is successfully loaded onto the quantum memories. The probability of generation is calculated simply by the trace of the un-normalized quantum state of the source in the specific instance of interest. (2) $F$, \textit{the fidelity} which we assume is the Bell state fidelity, which mathematically is
\begin{align}
    F = \frac{\bra{\xi} \rho \ket{\xi}}{\text{Tr}(\rho)}
\end{align}

Where $\ket{\xi}$ is the Bell state of interest. The Bell state produced by the cascaded source will be dependent upon the heralding click pattern. For example, if the source heralds the pattern $\ket{1,1,0,0}_{3,4,5,6}$ (in the mode convention of figure \ref{fig:diagram}) then the Bell state produced is the $\ket{\psi^-}$ state. 

\subsubsection{Physical Modeling Parameters}

The central parameter of interest for modeling entanglement sources is the mean photon number per mode, $\mu$, which is described in the previous subsection in terms of parameters such as the squeezing strengths. 


We consider physical non-idealities, including: 
\GR{
\begin{itemize}
    \item $\eta_b$, \textit{source heralding photon loss} which can occur due to lossy optical elements in the heralding process (i.e. beam splitters) 
    \item $\eta_t$, \textit{loss in the transmission of the state} which can occur because of optical fiber loss or free space transmission, depending on the network architecture
    \item $\eta_d$, \textit{detector specific loss} which can occur because of inefficient detectors used in the source heralding as well as receiver heralding
    \item $P_d$, \textit{dark click probability} which is the probability that our noisy photon detector registers a click despite a photon from the source not being present. Dark click probability can be used to model generic detector inefficiencies, as well as background noise. 
\end{itemize}
}

\section{Mathematical Principles}
\label{sec:math-overview}
\GR{\texttt{This entire section was rewritten, replacing the previous section with a summary of the mathematical methods. The previous version was moved to an appendix, and is referenced in this section.} }

In this section we summarize the mathematical model used in \texttt{genqo} to describe single-frequency-mode SPDC and cascaded/ZALM entanglement sources, and their loading into Duan-Kimble style quantum memories. All details of the derivations are collected in Appendix~\ref{app:math-details}. The goal is to provide a conceptual overview and to clarify how the underlying physics is exposed through the software interface.

At a high level, the modeling pipeline implemented in \texttt{genqo} proceeds through the following stages:
\begin{enumerate}
  \item \textbf{Gaussian description of the photonic source.}  
  The source is first represented as Gaussian states, specified by covariance matrices in phase space.
  \item \textbf{Conversion to a coherent-state (K-function) representation.}  
  The Gaussian state is rewritten in the coherent-state basis, yielding a compact ``K-function'' representation that is convenient for subsequent non-Gaussian operations such as loss and single-photon detection.
  \item \textbf{Application of non-Gaussian operations.}  
  Channel loss, detector imperfections, and single-photon detection are incorporated as Kraus operators acting on the coherent-state representation. The resulting integrals are evaluated using Gaussian-integration techniques (Wick's theorem or, equivalently, hafnians).
  \item \textbf{Extraction of figures of merit.}  
  From the resulting (generally mixed) photonic density matrix we compute the probability of success and the fidelity with respect to an ideal Bell state.
  \item \textbf{Mapping to a spin-spin state via quantum memories.}  
  Finally, we model loading into Duan-Kimble style memories, treating the memory qubits as bosonic dual-rail modes coupled to the photonic state via an effective cross-Kerr interaction. This yields the spin-spin density matrix that is relevant for higher-layer networking protocols.
\end{enumerate}

Throughout this section we highlight the corresponding \texttt{genqo} routines so that users can easily relate the mathematics to the software API.

\subsection{Gaussian description of SPDC and cascaded/ZALM sources}

The starting point is a Gaussian description of the photonic state emitted by the source. For each choice of physical parameters (mean photon number, losses, etc.), \texttt{genqo} constructs a covariance matrix
\begin{equation}
  V = V(\mu) \in \mathbb{R}^{2N \times 2N},
\end{equation}
for an $N$-mode system in terms of canonical quadrature operators
\begin{equation}
  \hat{x} = (\hat{q}_1, \hat{p}_1, \dots, \hat{q}_N, \hat{p}_N)^{\mathsf{T}}.
\end{equation}
The covariance matrix is defined in the usual way,
\begin{align}
    V_{ij} &= \frac{1}{2} \mmnt{ \{ \Delta \hat{x}_i , \Delta \hat{x}_j \} }\\
    \Delta \hat{x}_i &= \hat{x}_i - \mmnt{\hat{x}_i} \\
    \{ \hat{A}, \hat{B} \} &= \hat{A} \hat{B} + \hat{B} \hat{A}
\end{align}
with zero first moments for the states considered here.

The SPDC source is modeled as two, two-mode squeezed vacuum (TMSV) states combined via a mode swap, giving a dual-rail Bell-pair source. The cascaded/ZALM source is obtained by augmenting this configuration with a Bell-state measurement between idler modes from two SPDC sources. At the Gaussian level, all of these operations (beam splitters, permutations, etc.) correspond to symplectic transformations
\begin{equation}
  V \;\mapsto\; S V S^{\mathsf{T}},
\end{equation}
for an appropriate symplectic matrix $S$ built from elementary building blocks (beam splitters and permutations).

In the \texttt{genqo} implementation, this stage is encapsulated in the routine
\begin{center}
  \verb|calculate_covariance_matrix|
\end{center}
which returns the covariance matrix $V(\mu)$ for a given source class (SPDC or ZALM) and mean photon number $\mu$, after all required Gaussian transformations have been applied. This Gaussian description is the unique input to the subsequent, more general, hybrid Gaussian/non-Gaussian treatment.

\subsection{Hybrid Gaussian / non-Gaussian representation via the K-function}

The Gaussian state is then mapped into a coherent-state representation that is convenient for incorporating loss and detection. For an $N$-mode state, we expand in the coherent-state basis $\{ \lvert \boldsymbol{\alpha} \rangle \}$ with
\begin{equation}
  \boldsymbol{\alpha} = (\alpha_1, \ldots, \alpha_N) \in \mathbb{C}^N.
\end{equation}
We write the state as
\begin{equation}
  \lvert \psi \rangle = \int \mathrm{d}^{2N}\boldsymbol{\alpha}\,
  K(\boldsymbol{\alpha}) \lvert \boldsymbol{\alpha} \rangle,
\end{equation}
where $K(\boldsymbol{\alpha})$ is the so-called K-function.\footnote{For a zero-mean Gaussian state, $K(\boldsymbol{\alpha})$ has a closed-form expression entirely in terms of the covariance matrix $V$, as detailed in Appendix~\ref{app:k-function}.}

This representation plays two roles:
\begin{itemize}
  \item it provides a bridge between the compact Gaussian description ($V$) and density matrices; and
  \item it allows non-Gaussian elements (loss, detection, cross-Kerr interactions with memories) to be expressed as simple transformations of coherent amplitudes followed by Gaussian integrals.
\end{itemize}

In \texttt{genqo}, the construction of the K-function and its associated matrices is handled by
\begin{center}
  \verb|calculate_k_function_matrix|,
\end{center}
which takes $V$ as input and outputs the objects needed to evaluate density matrices and observables in the coherent-state picture. These objects, which combined we call $\mathbf{A}$ matrices, are derived in detail in Appendix \ref{sec:detection}.

\subsection{Modeling loss and single-photon detection}

Losses in the channels and detectors are incorporated at the level of the coherent-state representation using Kraus operators for amplitude damping. Physically relevant efficiencies include:
\begin{itemize}
  \item $\eta_{\mathrm{b}}$ (heralding/Bell-state measurement efficiency),
  \item $\eta_{\mathrm{t}}$ (transmission or out-coupling efficiency), and
  \item $\eta_{\mathrm{d}}$ (detector efficiency),
\end{itemize}
along with a dark-click probability $P_{\mathrm{d}}$ that models detector dark noise, i.e. spurious detector clicks from unavoidable noise sources.

Mathematically, these elements act as transformations of the coherent amplitudes in the K-function. For example, a pure-loss channel with transmissivity $\eta$ maps
\begin{equation}
  \alpha \;\mapsto\; \sqrt{\eta}\,\alpha
\end{equation}
in each affected mode, with the ``lost'' part traced over. Detector dark counts are included by summing over alternative click patterns weighted by appropriate classical probabilities $P_{\mathrm{d}}$.

Single-photon detection is implemented as projection onto Fock states in the coherent-state representation. The resulting expressions involve Gaussian integrals of polynomials in the coherent amplitudes (or quadrature variables). We call these polynomials $C$ coefficients, which are derived in Appendix \ref{sec:detection}. Together with the $\mathbf{A}$ matrices, these Gaussian integrals are evaluated using Wick's theorem (or, equivalently, hafnians of appropriate matrices), which reduces high-order moments to sums over pairings determined by the photon click pattern. In the code this machinery is organized around an intermediate matrix $A$ (and its modified versions) and a helper function
\begin{center}
  \verb|W(A, C)|,
\end{center}
which computes the relevant Gaussian moments. The detailed structure of $A$ and the Wick/hafnian calculations is given in Appendix~\ref{app:wick}.

\subsection{Figures of merit and physical parameters}

The key performance parameters returned by \texttt{genqo} are:
\begin{enumerate}
  \item The \emph{probability of successful generation} $P_{\mathrm{gen}}$, i.e., the probability that the specified heralding pattern is observed at the source (and, when relevant, that subsequent memory-loading heralding detections occur); and
  \item The \emph{Bell-state fidelity} $F_{\mathrm{Bell}}$ of the resulting state with respect to a target Bell state.
\end{enumerate}

For a photonic state $\rho_{\mathrm{ph}}$ after application of losses and detection, and a target dual-rail Bell state $\lvert \Phi_{\mathrm{Bell}} \rangle$, we define
\begin{align}
  P_{\mathrm{gen}}
  &= \mathrm{Tr}[\rho_{\mathrm{ph}}], \\
  F_{\mathrm{Bell}}
  &= \frac{ \langle \Phi_{\mathrm{Bell}} \lvert \rho_{\mathrm{ph}} \rvert \Phi_{\mathrm{Bell}} \rangle }
          { P_{\mathrm{gen}} }.
\end{align}
Here $\rho_{\mathrm{ph}}$ is generally an unnormalized post-selected state conditioned on the desired detection pattern.

In practice, both quantities are functions of the mean photon number $\mu$ and the various efficiencies,
\begin{equation}
  P_{\mathrm{gen}} = P_{\mathrm{gen}}(\mu, \eta_{\mathrm{b}}, \eta_{\mathrm{t}},
  \eta_{\mathrm{d}}, p_{\mathrm{dc}}),
\end{equation}
and similarly for $F_{\mathrm{Bell}}$. The core \texttt{genqo} routines
\begin{center}
  \verb|calculate_probability_success| \quad \text{and} \quad \verb|calculate_fidelity|
\end{center}
implement these expressions in terms of the K-function matrices, the loss parameters, and the Wick/hafnian machinery. The detailed formulas are given in Appendix~\ref{app:prob-fid}.

\subsection{Spin-spin state via Duan-Kimble quantum memories}

The final stage of the model considers the state of two quantum memories after loading the photonic Bell pair produced by the source. We follow the Duan-Kimble protocol, in which a dual-rail photonic qubit is mapped onto an atomic (or solid-state) qubit using a sequence of operations:
\begin{enumerate}
  \item An effective controlled-phase (CZ) interaction between the photonic mode and the memory qubit,
  \item Interference of photonic modes on beam splitters (e.g., a polarizing beam splitter for polarization-encoded dual-rail qubits), and
  \item Single-photon detection, which heralds successful loading.
\end{enumerate}

In our formalism, the memory qubits are modeled as additional bosonic modes prepared in an appropriate initial state. The CZ interaction is implemented as a cross-Kerr unitary between the photonic and memory modes, and the subsequent interference and detection are treated using the same K-function and Wick/hafnian machinery used for the source itself. Click patterns of the photonic modes yields a spin-spin density matrix
\begin{equation}
  \rho_{\mathrm{spin}} = \rho_{\mathrm{spin}}(\mu, \eta_{\mathrm{b}},
  \eta_{\mathrm{t}}, \eta_{\mathrm{d}}, p_{\mathrm{dc}}),
\end{equation}
which can be analyzed using the same figures-of-merit as the purely photonic state (e.g., Bell-state fidelity, concurrence, etc.).

The corresponding implementation in \texttt{genqo} reuses the same matrices and moment-calculation routines as in the purely photonic case; from the user’s perspective, the memory loading is simply a different configuration of detection patterns and loss parameters. The full derivation of the memory-loading map and the expression for $\rho_{\mathrm{spin}}$ is presented in Appendix~\ref{app:memory}.

Figure \ref{fig:block} describes the general flow the quantities described in this section are used within \texttt{genqo}. Further details on this calculation can be found in Appendix \ref{app:math-details}.

\begin{figure}[h]
    \centering
    \includegraphics[width=\linewidth]{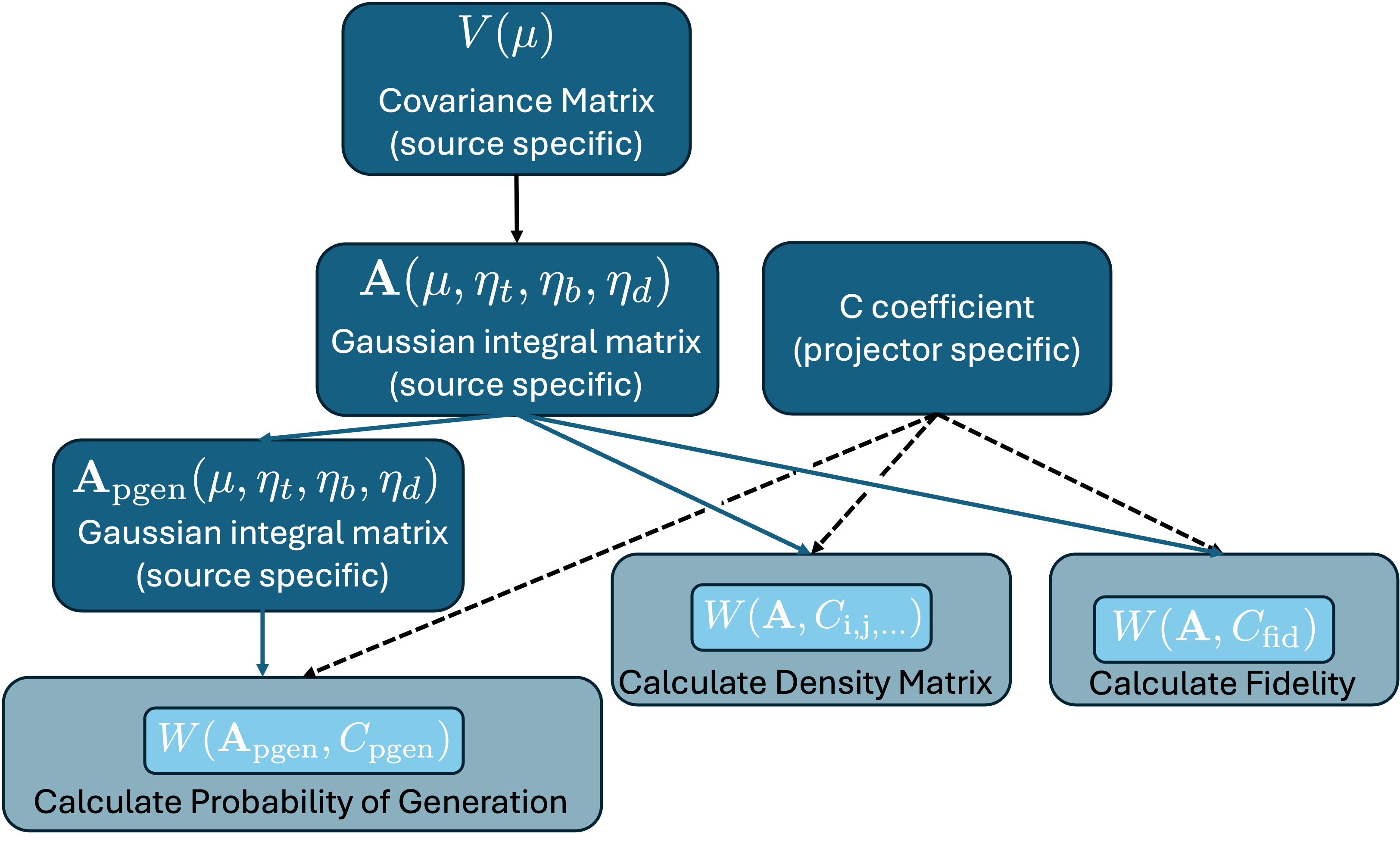}
    \caption{\GR{A block diagram of the flow of a hybrid Gaussian/non-Gaussian source calculation. The calculation begins by defining the covariance matrix $V$, that can be used to calculate the $\mathbf{A}$ matrix, whose elements define the solutions for Gaussian integrals of two quadrature variables. Using $\mathbf{A}$ together with a $C$ coefficient, which is defined by a projection operation, the $W$ function is used to calculate quantities of interest such as the density matrix and the fidelity. The $\mathbf{A}$ matrix can also be extended to construct $\mathbf{A_{\text{pgen}}}$, which is similarly used with a unique $C$ coefficient to calculate the probability of generation $P_{\text{gen}}$. $\mathbf{A_{\text{pgen}}}$ is unique to $\mathbf{A}$ because of the trace operation required to calculate $P_{\text{gen}}$, as compared to the projection operations for the density matrix and fidelity calculation. A detailed derivation of these matrices and coefficients can be found in the appendix, especially \ref{app:prob-fid} and \ref{sec:detection} } }
    \label{fig:block}
\end{figure}

\section{Simulation Module}

This work focuses on providing a highly accurate model of the ZALM cascaded entanglement source, and for the sake of reusability and reproducibility, a stack of open source software tools implementing the presented modeling methods is provided. These tools come with significant quality assurance measures, including proper continuous integration development processes, testing and verification suites, and ease of installation through popular software package indexes. In this section we discuss these tools, examining the low-level Python implementation of these bespoke modeling techniques (the \texttt{genqo} Python package), its integration as one of the models available in the QuantumSymbolics.jl Julia computer algebra system, and its use in the full-stack networking simulator QuantumSavory.jl. We present the low-level ZALM-specific Python API, the visulization and exploration tools for studying the parameterization of the ZALM source among others within QuantumSavory, and examples of full-stack simulations. Lastly, as prescribed in the "Software for Quantum Networks" special issue, we provide a stand-alone "mock hardware" HTTP server which can be used as a universal API for this software, in addition to the already existing APIs.

\subsection{The genqo low-level modeling tool}

\begin{figure}[h]
    \centering
    \includegraphics[width=\linewidth]{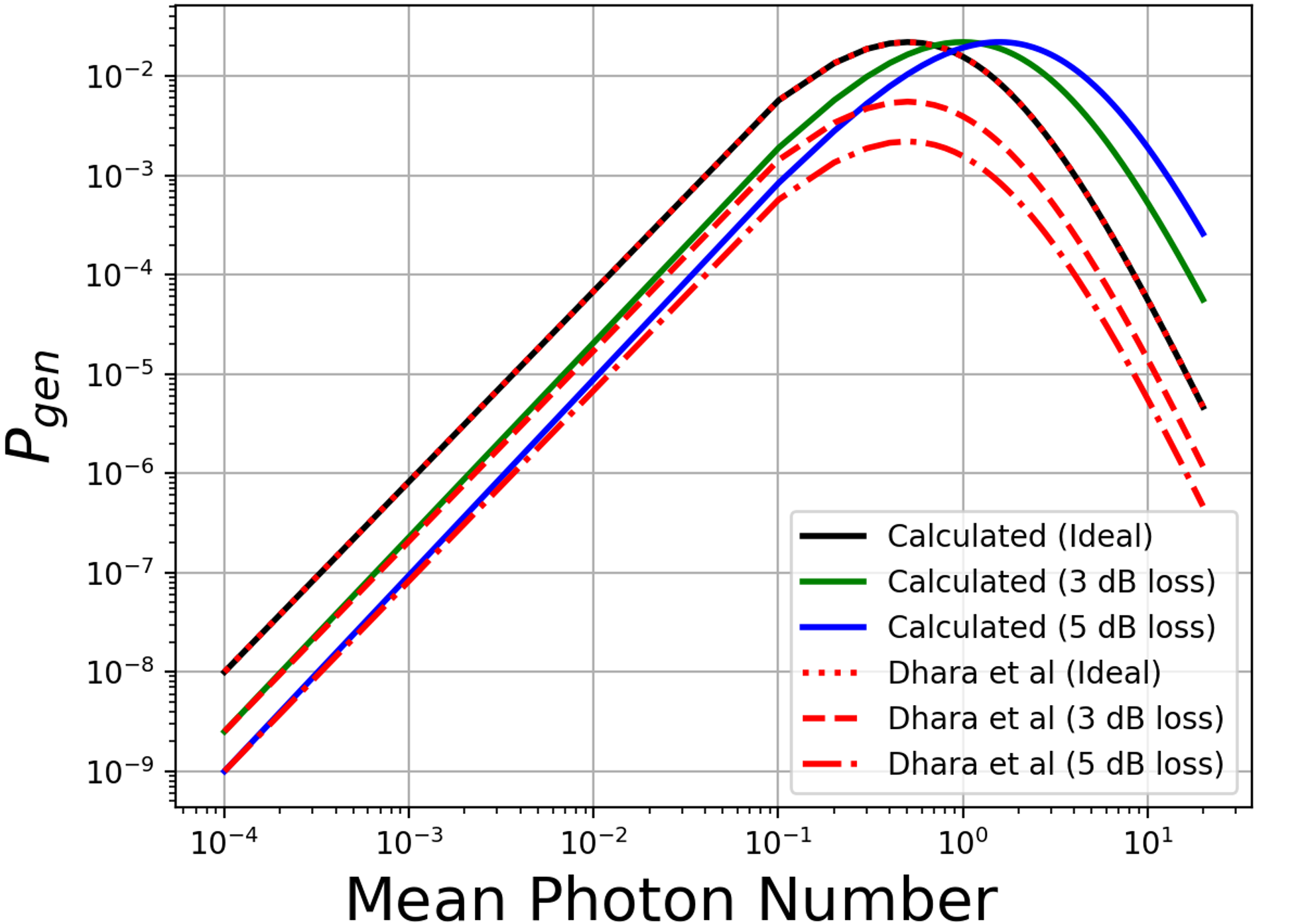}
    \caption{The probability of successful heralded pair generation versus the mean photon number, calculated using both our hybrid Gaussian/non-Gaussian model (black/green/blue) as well as the older low-mean-photon-number-approximation (red). An important property uncovered by our more precise model is that the peak generation probability does not plummet with increased loss -- one simply needs a high mean photon number to compensate. Naturally, the fidelity of entanglement will drop at higher mean photon numbers due to multi-photon events, nonetheless this discovery opens up exciting future prospects in employing distillation to achieve unexpectedly high rates of generation at good fidelities.}
    \label{fig:gaussvlowns}
\end{figure}

\texttt{genqo} is a Python library implementing the modeling formalism described in previous sections for SPDC and ZALM entanglement sources. It is a standalone tool with minimal dependencies that can be install\GR{ed} from the Python package index on any platform. Below we showcase the API, with examples of parameterizing an entanglement source and evaluating the success probability versus the mean photon number for such a source. The results are given in 
Figure \ref{fig:gaussvlowns}, demonstrating a discrepancy between the earlier approximate models (as presented in Dhara et al~\cite{dhara_heralded_2022}) and our more complete results.

\begin{widetext}
\begin{lstlisting}[language=Python]
    import numpy as np
    from genqo import ZALM
    
    Pgenv = np.array([]) # Define an array to store the probability of generaiton 
    muv = np.linspace(10**(-4),20,200) # The mean photon number values over which we will sweep
    zalm_example = ZALM()
    zalm_example.params["bsm_efficiency"] = 1 # No loss in the BSM
    # zalm_example.params["bsm_efficiency"] = 10**(-3/10) # or, e.g. 3 dB of loss
    zalm_example.params["outcoupling_efficiency"] = 1 # No loss in the transmission
    zalm_example.params["detection_efficiency"] = 1 # No loss in any of the detectors
    for i in muv:
        zalm_example.params["mean_photon"] = i
        zalm_example.run()
        zalm_example.calculate_probability_success()
        Pgenv = np.append(Pgenv, zalm_example.results["probability_success"])
\end{lstlisting}
\end{widetext}

In particular, we observe that in the presence of loss, the approximated approach expects the probability of generation to peak at a single value of the mean photon number $\mu$, with the peak plummeting as losses increase. On the other hand, using our complete Gaussian model we observe the peak shifting to higher mean photon number values for higher loss, retaining a high success probability. This discrepancy opens up new opportunities: while higher mean photon number implies also lowered fidelity, we can apply entanglement distillation to explore a rate-fidelity tradeoff that would have been unavailable to us according to the earlier approximate models of the ZALM source. The full-stack simulation tools described below, incorporating the genqo models, make such a study possible. 

A jupyter notebook with examples is found in the genqo \GR{Zenodo repository, "tutorial" folder ~\cite{JGRichardson_StefanKrastanov_2025}, acting as a long-term archival link. This notebook can facilitate users of genqo without requiring prior knowledge of covariance matrix formalism.} 

\subsection{QuantumSymbolics.jl computer algebra system}

The \texttt{genqo} model is integrated into QuantumSymbolics.jl, a computer algebra system (CAS) for quantum mechanics and quantum information. QuantumSymbolics.jl allows for the algebraic manipulation of quantum states and operators in a symbolic representation. A key feature of this CAS is its {\tt express} API, which translates these symbolic objects into various numerical representations suitable for different simulation backends, such as state vectors, density matrices, or stabilizer tableaux.

Within this framework, the ZALM source is represented by the {\tt GenqoMultiplexedCascadedBellPairW} struct. An instance of this struct, such as {\tt s = GenqoMultiplexedCascadedBellPairW($\eta^b$=1.0, $\eta^d$=1.0, $\eta^t$=0.8, $\mu$=0.1, $P^d$=1e-8)}, is a symbolic object that encapsulates the full, highly-parameterized physical model of the entanglement source. This object can be integrated into larger symbolic expressions representing more complex quantum information processing tasks.

Symbolic operations can be performed on these \texttt{genqo}-backed states before any numerical evaluation is triggered. For instance, to model the state after transmission through a lossy channel, one can define a symbolic representation of that channel {\tt L}, and apply it to the source object: {\tt s\_final = L(0.5) * s}. This composition is handled purely at the symbolic level. The expensive numerical computation, which involves calling the \texttt{genqo} Python backend to generate the source's density matrix, is deferred until the {\tt express(s\_final)} command is issued. At that point, QuantumSymbolics.jl resolves the entire expression: it first calls \texttt{genqo} to get the initial numerical state for the given parameters, and then applies a numerical representation of the loss channel to that state, caching intermediary results. This methodology allows for the seamless composition of a bespoke, high-accuracy model like \texttt{genqo} with the extensive library of general-purpose symbolic quantum operations available in QuantumSymbolics.jl, facilitating sophisticated, multi-component system modeling. Moreover, this approach permits symbolic objects to be expressed in a variety of different formalisms -- the {\tt express} function can convert not only to a state vector numerical object, but also to Gaussian quantum states, stabilizer states, or tensor networks, whenever feasible.

\subsection{QuantumSavory.jl full-stack network simulator}

The \texttt{genqo} models, being fully integrated into QuantumSymbolics.jl, are immediately available for use within QuantumSavory.jl, a full-stack, discrete-event simulator for quantum networks. QuantumSavory.jl is designed to model the complex interplay of quantum state evolution, classical control messaging, and network topology, making it an ideal platform for assessing the system-level performance of quantum hardware. Because the simulator uses QuantumSymbolics.jl as its frontend for describing arbitrary quantum dynamics, any symbolic object, including the {\tt GenqoMultiplexedCascadedBellPairW} state, can be seamlessly initialized in a network's quantum registers.

To demonstrate this capability, we have developed a simulation of a first-generation quantum repeater chain. In this scenario, entanglement is first established between adjacent nodes of the network and then extended end-to-end via entanglement swapping. The simulation is parameterized by the choice of the initial entanglement source. By simply switching the state initialized at the source nodes -- for example, from an idealized perfect Bell pair to the physically realistic {\tt GenqoMultiplexedCascadedBellPairW} -- a user can directly compare the performance of different hardware assumptions. The simulator tracks figures of merit such as end-to-end fidelity and entanglement generation rates, allowing for a direct assessment of how the non-ideal properties of the \texttt{genqo} source, such as multi-photon emission noise at high pump powers, impact the overall network performance. This enables a comprehensive, system-level analysis of the rate-versus-fidelity tradeoffs uncovered by the \texttt{genqo} model.

An interactive version of this simulation, allowing real-time parameter exploration, is available at \url{areweentangledyet.com/simplerepeaters/}.

\subsection{State Explorer visualization tool}

To facilitate an intuitive understanding of the complex parameter space of the \texttt{genqo} sources and other sources available in QuantumSavory.jl, we provide an interactive visualization tool, the "State Explorer". This web-based application, built on the Makie plotting library and integrated into QuantumSavory, allows for the real-time exploration of any two-qubit state, including the {\tt GenqoMultiplexedCascadedBellPairW}.

Users can manipulate sliders corresponding to the source's physical parameters, such as mean photon number and various coupling efficiencies. The tool dynamically recalculates and displays crucial figures of merit along multiple parameter dimensions, including generation probability and fidelity with respect to a perfect Bell pair. This provides a direct, visual confirmation of the trade-offs discussed in this work, such as the ability to recover high success probabilities at the cost of fidelity by increasing the pump power in lossy regimes.

The State Explorer is available within the QuantumSavory code base and a live instance is provided at \url{areweentangledyet.com/state_explorer/}.

\subsection{JSON Server}

To ensure the broadest possible accessibility and interoperability of our models, and in accordance with the guidelines for this special issue, we provide a standalone HTTP server that exposes the \texttt{genqo} model and other states available in QuantumSavory.jl through a universal, language-agnostic API. This server acts as a "mock hardware" interface, accepting requests with specified state parameters and state representation. Upon receiving a request, the server can provide the full density matrix of the requested state, its success probability, and other relevant figures of merit like fidelity and concurrence, all formatted in a standard data-interchange format like JSON. This allows researchers and developers to integrate our high-accuracy models into any third-party simulation or analysis tool capable of making standard web requests, without needing to directly engage with the underlying Julia or Python codebases. The tool is available at \url{areweentangledyet.com/states_rest_server/docs}.

\section{Discussion and Conclusions}

The ZALM source modeled in this work is a practical, deployable entanglement source overcoming multiplexing and heralding issues plaguing earlier blueprints. We provide a digital twin of such a source at an accuracy significantly surpassing previous approximate results, particularly in the high-photon-number regime, where new interesting tradeoffs are uncovered. Our model shows that drops in probability of generation due to loss can reliably be circumvented by increasing the mean photon number, a capability that was not uncovered in earlier models. While this newly interesting parameter regime suffers from lower entanglement fidelity, it poses interesting opportunities for entanglement distillation to be employed on top of a ZALM source to achieve unexpectedly high overall rates and fidelities -- a prospect that deserves further exploration. Such exploration would be much easier thanks to the detailed, reproducible, tested, easy-to-install software models we provide.

The modular "polyglot" way in which a bespoke high-quality model (\texttt{genqo}) implemented in an arbitrary language is made seamlessly available in a holistic computer algebra system (QuantumSymbolics.jl) and incorporated in a full-stack quantum network simulator (QuantumSavory.jl) bodes well for the future of the modeling software ecosystem. It is a proof that the diverse tools developed in disconnected research labs can be brought together. Modern development practices (heavy use of package and dependency management, automated testing, static analysis, and continuous integration on all popular hardware platforms and operating systems) turn what used to be a difficult technical problem into a much simpler social problem of coordination, collaboration, and maintaining open access to results. What used to be a significant standardization effort and overhead of supporting such standardization is now a much simpler task thanks to the existence of engineering tools that interface different languages and numerical storage formats as well as scientific tools that convert between different formalisms (e.g. state vectors, Gaussian quantum formalism, stabilizer states, tensor networks, and more).


\section*{Acknowledgements}
We thank Yousef Chahine for valuable discussions and insights related to this work. This work was supported by a NASA Space Technology Graduate Research Opportunity. We gratefully acknowledge support from the NSF Engineering Research Center for Quantum Networks (Grant No. EEC-1941583), from NSF grant 2346089, and from NSF grant 2402861. 
\bibliography{bibliography}

@article{dhara_heralded_2022,
	title = {Heralded {Multiplexed} {High}-{Efficiency} {Cascaded} {Source} of {Dual}-{Rail} {Entangled} {Photon} {Pairs} {Using} {Spontaneous} {Parametric} {Down}-{Conversion}},
	volume = {17},
	issn = {2331-7019},
	url = {https://link.aps.org/doi/10.1103/PhysRevApplied.17.034071},
	doi = {10.1103/PhysRevApplied.17.034071},
	 
	number = {3},
	urldate = {2023-05-26},
	journal = {Physical Review Applied},
	author = {Dhara, Prajit and Johnson, Spencer J. and Gagatsos, Christos N. and Kwiat, Paul G. and Guha, Saikat},
	month = mar,
	year = {2022},
	pages = {034071},
}

@article{chen_zero-added-loss_2023,
	title = {Zero-{Added}-{Loss} {Entangled}-{Photon} {Multiplexing} for {Ground}- and {Space}-{Based} {Quantum} {Networks}},
	volume = {19},
	issn = {2331-7019},
	url = {https://link.aps.org/doi/10.1103/PhysRevApplied.19.054029},
	doi = {10.1103/PhysRevApplied.19.054029},
	 
	number = {5},
	urldate = {2023-05-26},
	journal = {Physical Review Applied},
	author = {Chen, Kevin C. and Dhara, Prajit and Heuck, Mikkel and Lee, Yuan and Dai, Wenhan and Guha, Saikat and Englund, Dirk},
	month = may,
	year = {2023},
	pages = {054029},
}

@article{schneeloch_introduction_2019,
	title = {Introduction to the absolute brightness and number statistics in spontaneous parametric down-conversion},
	volume = {21},
	issn = {2040-8978, 2040-8986},
	url = {https://iopscience.iop.org/article/10.1088/2040-8986/ab05a8},
	doi = {10.1088/2040-8986/ab05a8},
	 
	number = {4},
	urldate = {2023-11-28},
	journal = {J. Opt.},
	author = {Schneeloch, James and Knarr, Samuel H and Bogorin, Daniela F and Levangie, Mackenzie L and Tison, Christopher C and Frank, Rebecca and Howland, Gregory A and Fanto, Michael L and Alsing, Paul M},
	month = apr,
	year = {2019},
	pages = {043501},
}

@article{grice_spectral_1997,
	title = {Spectral information and distinguishability in type-{II} down-conversion with a broadband pump},
	volume = {56},
	issn = {1050-2947, 1094-1622},
	url = {https://link.aps.org/doi/10.1103/PhysRevA.56.1627},
	doi = {10.1103/PhysRevA.56.1627},
	 
	number = {2},
	urldate = {2023-09-22},
	journal = {Phys. Rev. A},
	author = {Grice, W. P. and Walmsley, I. A.},
	month = aug,
	year = {1997},
	pages = {1627--1634},
}

@misc{lvovsky_squeezed_2016,
	title = {Squeezed light},
	url = {http://arxiv.org/abs/1401.4118},
	urldate = {2023-11-29},
	publisher = {arXiv},
	author = {Lvovsky, A. I.},
	month = jul,
	year = {2016},
	note = {arXiv:1401.4118 [physics, physics:quant-ph]
version: 2},
}

@article{khabiboulline2019optical,
  title={Optical interferometry with quantum networks},
  author={Khabiboulline, Emil T and Borregaard, Johannes and De Greve, Kristiaan and Lukin, Mikhail D},
  journal={Physical review letters},
  volume={123},
  number={7},
  pages={070504},
  year={2019},
  publisher={APS}
}

@article{chen_polarization_2021,
	title = {A polarization encoded photon-to-spin interface},
	volume = {7},
	issn = {2056-6387},
	url = {https://www.nature.com/articles/s41534-020-00337-3},
	doi = {10.1038/s41534-020-00337-3},
	 
	number = {1},
	urldate = {2023-05-26},
	journal = {npj Quantum Information},
	author = {Chen, K. C. and Bersin, E. and Englund, D.},
	month = jan,
	year = {2021},
	pages = {2},
}

@article{weedbrook_gaussian_2012,
	title = {Gaussian {Quantum} {Information}},
	volume = {84},
	issn = {0034-6861, 1539-0756},
	url = {http://arxiv.org/abs/1110.3234},
	doi = {10.1103/RevModPhys.84.621},
	number = {2},
	urldate = {2024-04-01},
	journal = {Rev. Mod. Phys.},
	author = {Weedbrook, Christian and Pirandola, Stefano and Garcia-Patron, Raul and Cerf, Nicolas J. and Ralph, Timothy C. and Shapiro, Jeffrey H. and Lloyd, Seth},
	month = may,
	year = {2012},
	note = {arXiv:1110.3234 [quant-ph]},
	pages = {621--669},
}

@article{davis_improved_2022,
	title = {Improved heralded single-photon source with a photon-number-resolving superconducting nanowire detector},
	volume = {18},
	issn = {2331-7019},
	url = {http://arxiv.org/abs/2112.11430},
	doi = {10.1103/PhysRevApplied.18.064007},
	number = {6},
	urldate = {2024-04-04},
	journal = {Phys. Rev. Applied},
	author = {Davis, Samantha I. and Mueller, Andrew and Valivarthi, Raju and Lauk, Nikolai and Narvaez, Lautaro and Korzh, Boris and Beyer, Andrew D. and Colangelo, Marco and Berggren, Karl K. and Shaw, Matthew D. and Sinclair, Neil and Spiropulu, Maria},
	month = dec,
	year = {2022},
	note = {arXiv:2112.11430 [physics, physics:quant-ph]},
	pages = {064007},
}

@article{gagatsos_efficient_2019,
	title = {Efficient representation of {Gaussian} states for multi-mode non-{Gaussian} quantum state engineering via subtraction of arbitrary number of photons},
	volume = {99},
	issn = {2469-9926, 2469-9934},
	url = {http://arxiv.org/abs/1902.01460},
	doi = {10.1103/PhysRevA.99.053816},
	number = {5},
	urldate = {2024-04-04},
	journal = {Phys. Rev. A},
	author = {Gagatsos, Christos and Guha, Saikat},
	month = may,
	year = {2019},
	note = {arXiv:1902.01460 [quant-ph]},
	pages = {053816},
}

@article{takeoka_full_2015,
	title = {Full analysis of multi-photon pair effects in spontaneous parametric down conversion based photonic quantum information processing},
	volume = {17},
	issn = {1367-2630},
	url = {https://dx.doi.org/10.1088/1367-2630/17/4/043030},
	doi = {10.1088/1367-2630/17/4/043030},
	 
	number = {4},
	urldate = {2024-04-08},
	journal = {New J. Phys.},
	author = {Takeoka, Masahiro and Jin, Rui-Bo and Sasaki, Masahide},
	month = apr,
	year = {2015},
	note = {Publisher: IOP Publishing},
	pages = {043030},
}

@article{duan_scalable_2004,
	title = {Scalable {Photonic} {Quantum} {Computation} through {Cavity}-{Assisted} {Interactions}},
	volume = {92},
	url = {https://link.aps.org/doi/10.1103/PhysRevLett.92.127902},
	doi = {10.1103/PhysRevLett.92.127902},
	number = {12},
	urldate = {2023-06-24},
	journal = {Physical Review Letters},
	author = {Duan, L.-M. and Kimble, H. J.},
	month = mar,
	year = {2004},
	note = {Publisher: American Physical Society},
	pages = {127902},
}

@article{khabiboulline_quantum-assisted_2019,
	title = {Quantum-assisted telescope arrays},
	volume = {100},
	url = {https://link.aps.org/doi/10.1103/PhysRevA.100.022316},
	doi = {10.1103/PhysRevA.100.022316},
	number = {2},
	urldate = {2023-06-24},
	journal = {Phys. Rev. A},
	author = {Khabiboulline, E. T. and Borregaard, J. and De Greve, K. and Lukin, M. D.},
	month = aug,
	year = {2019},
	note = {Publisher: American Physical Society},
	pages = {022316},
}

@article{kwiat_ultrabright_1999,
	title = {Ultrabright source of polarization-entangled photons},
	volume = {60},
	issn = {1050-2947, 1094-1622},
	url = {https://link.aps.org/doi/10.1103/PhysRevA.60.R773},
	doi = {10.1103/PhysRevA.60.R773},
	 
	number = {2},
	urldate = {2023-11-28},
	journal = {Phys. Rev. A},
	author = {Kwiat, Paul G. and Waks, Edo and White, Andrew G. and Appelbaum, Ian and Eberhard, Philippe H.},
	month = aug,
	year = {1999},
	pages = {R773--R776},
}

@article{kwiat_new_1995,
	title = {New {High}-{Intensity} {Source} of {Polarization}-{Entangled} {Photon} {Pairs}},
	volume = {75},
	issn = {0031-9007, 1079-7114},
	url = {https://link.aps.org/doi/10.1103/PhysRevLett.75.4337},
	doi = {10.1103/PhysRevLett.75.4337},
	 
	number = {24},
	urldate = {2023-11-28},
	journal = {Phys. Rev. Lett.},
	author = {Kwiat, Paul G. and Mattle, Klaus and Weinfurter, Harald and Zeilinger, Anton and Sergienko, Alexander V. and Shih, Yanhua},
	month = dec,
	year = {1995},
	pages = {4337--4341},
}

@article{shapiro_entanglement_2024,
	title = {Entanglement source and quantum memory analysis for zero-added-loss multiplexing},
	volume = {22},
	url = {https://link.aps.org/doi/10.1103/PhysRevApplied.22.044014},
	doi = {10.1103/PhysRevApplied.22.044014},
	number = {4},
	urldate = {2024-11-05},
	journal = {Phys. Rev. Appl.},
	author = {Shapiro, Jeffrey H. and Raymer, Michael G. and Embleton, Clark and Wong, Franco N.C. and Smith, Brian J.},
	month = oct,
	year = {2024},
	note = {Publisher: American Physical Society},
	pages = {044014},
}

@article{gottesman_longer-baseline_2012,
    title = {Longer-{Baseline} {Telescopes} {Using} {Quantum} {Repeaters}},
    volume = {109},
    url = {https://link.aps.org/doi/10.1103/PhysRevLett.109.070503},
    doi = {10.1103/PhysRevLett.109.070503},
    number = {7},
    urldate = {2024-09-06},
    journal = {Physical Review Letters},
    author = {Gottesman, Daniel and Jennewein, Thomas and Croke, Sarah},
    month = aug,
    year = {2012},
    note = {Publisher: American Physical Society},
    pages = {070503},
}

@article{van_milligen_utilizing_2024,
    title = {Utilizing probabilistic entanglement between sensors in quantum networks},
    volume = {22},
    url = {https://link.aps.org/doi/10.1103/PhysRevApplied.22.064085},
    doi = {10.1103/PhysRevApplied.22.064085},
    number = {6},
    urldate = {2025-06-20},
    journal = {Physical Review Applied},
    author = {Van Milligen, Emily A. and Gagatsos, Christos N. and Kaur, Eneet and Towsley, Don and Guha, Saikat},
    month = dec,
    year = {2024},
    note = {Publisher: American Physical Society},
    pages = {064085},
}

@misc{shapiro_high-fidelity_2025,
    title = {High-fidelity, quasi-deterministic entanglement generation using phase-matched spectral islands in a zero-added-loss multiplexing architecture},
    url = {http://arxiv.org/abs/2507.14427},
    doi = {10.48550/arXiv.2507.14427},
    urldate = {2025-07-22},
    publisher = {arXiv},
    author = {Shapiro, Jeffrey H. and Embleton, Clark and Raymer, Michael G. and Smith, Brian J.},
    month = jul,
    year = {2025},
    note = {arXiv:2507.14427 [quant-ph]},
}

@book{Ou_2007, address={New York}, title={Multi-photon Quantum interference}, ISBN={9780387255323}, callNumber={QC174.17.Q33 O9 2007}, publisher={Springer}, author={Ou, Zhe-Yu Jeff}, year={2007} }

@book{Boyd_2003, address={San Diego, CA}, edition={2nd ed.}, title={Nonlinear optics}, ISBN={9780080479750}, abstractNote={The Optical Society of America (OSA) and SPIE The International Society for Optical Engineering have awarded Robert Boyd with an honorable mention for the Joseph W. Goodman Book Writing Award for his work on Nonlinear Optics, 2nd edition. Nonlinear optics is essentially the study of the interaction of strong laser light with matter. It lies at the basis of the field of photonics, the use of light fields to control other light fields and to perform logical operations. Some of the topics of this book include the fundamentals and applications of optical systems based on the nonlinear interaction of light with matter. Topics to be treated include: mechanisms of optical nonlinearity, second-harmonic and sum- and difference-frequency generation, photonics and optical logic, optical self-action effects including self-focusing and optical soliton formation, optical phase conjugation, stimulated Brillouin and stimulated Raman scattering, and selection criteria of nonlinear optical materials. Covers all the latest topics and technology in this ever-evolving area of study that forms the backbone of the major applications of optical technology Offers first-rate instructive style making it ideal for self-study Emphasizes the fundamentals of non-linear optics rather than focus on particular applications that are constantly changing.}, publisher={Academic Press}, author={Boyd, Robert W.}, year={2003}}

@article{schumaker_new_1985,
    title = {New formalism for two-photon quantum optics. {II}. {Mathematical} foundation and compact notation},
    volume = {31},
    copyright = {http://link.aps.org/licenses/aps-default-license},
    issn = {0556-2791},
    url = {https://link.aps.org/doi/10.1103/PhysRevA.31.3093},
    doi = {10.1103/PhysRevA.31.3093},
    language = {en},
    number = {5},
    urldate = {2025-01-30},
    journal = {Physical Review A},
    author = {Schumaker, Bonny L. and Caves, Carlton M.},
    month = may,
    year = {1985},
    pages = {3093--3111},
}

@article{kim_phase-stable_2006,
    title = {Phase-stable source of polarization-entangled photons using a polarization {Sagnac} interferometer},
    volume = {73},
    url = {https://link.aps.org/doi/10.1103/PhysRevA.73.012316},
    doi = {10.1103/PhysRevA.73.012316},
    number = {1},
    urldate = {2025-07-21},
    journal = {Physical Review A},
    author = {Kim, Taehyun and Fiorentino, Marco and Wong, Franco N. C.},
    month = jan,
    year = {2006},
    note = {Publisher: American Physical Society},
    pages = {012316},
}

@article{wong_efficient_2006,
    title = {Efficient generation of polarization-entangled photons in a nonlinear crystal},
    volume = {16},
    issn = {1555-6611},
    url = {https://doi.org/10.1134/S1054660X06110053},
    doi = {10.1134/S1054660X06110053},
    language = {en},
    number = {11},
    urldate = {2025-09-23},
    journal = {Laser Physics},
    author = {Wong, F. N. C. and Shapiro, J. H. and Kim, T.},
    month = nov,
    year = {2006},
    pages = {1517--1524},
}

@article{zhang_distributed_2021,
    title = {Distributed quantum sensing},
    volume = {6},
    issn = {2058-9565},
    url = {https://dx.doi.org/10.1088/2058-9565/abd4c3},
    doi = {10.1088/2058-9565/abd4c3},
    language = {en},
    number = {4},
    urldate = {2025-09-23},
    journal = {Quantum Science and Technology},
    author = {Zhang, Zheshen and Zhuang, Quntao},
    month = jul,
    year = {2021},
    note = {Publisher: IOP Publishing},
    pages = {043001},
}

@article{fitzsimons_private_2017,
    title = {Private quantum computation: an introduction to blind quantum computing and related protocols},
    volume = {3},
    copyright = {2017 The Author(s)},
    issn = {2056-6387},
    shorttitle = {Private quantum computation},
    url = {https://www.nature.com/articles/s41534-017-0025-3},
    doi = {10.1038/s41534-017-0025-3},
    language = {en},
    number = {1},
    urldate = {2024-01-06},
    journal = {npj Quantum Information},
    author = {Fitzsimons, Joseph F.},
    month = jun,
    year = {2017},
    note = {Number: 1
Publisher: Nature Publishing Group},
    pages = {1--11},
}

@article{wehner_quantum_2018,
    title = {Quantum internet: {A} vision for the road ahead},
    volume = {362},
    shorttitle = {Quantum internet},
    url = {https://www.science.org/doi/10.1126/science.aam9288},
    doi = {10.1126/science.aam9288},
    number = {6412},
    urldate = {2025-09-23},
    journal = {Science},
    author = {Wehner, Stephanie and Elkouss, David and Hanson, Ronald},
    month = oct,
    year = {2018},
    note = {Publisher: American Association for the Advancement of Science},
    pages = {eaam9288},
}

@misc{Palkanoglou_2018, title={Complex Gaussian Integral}, url={https://math.stackexchange.com/questions/2667420/complex-gaussian-integral}, journal={Mathematics Stack Exchange}, author={Palkanoglou, George}, year={2018}, month=may }

@inbook{Zinn-Justin_2021, edition={5}, title={Gaussian integrals. Algebraic preliminaries}, ISBN={9780198834625}, url={https://academic.oup.com/book/43933/chapter/368824539}, DOI={10.1093/oso/9780198834625.003.0001}, abstractNote={Abstract              In this work, the perturbative aspects of quantum mechanics (QM) and quantum field theory (QFT), to a large extent, are studied with functional (path or field) integrals and functional techniques. This physics textbook thus begins with a discussion of algebraic properties of Gaussian measures, and Gaussian expectation values for a finite number of variables. The important role of Gaussian measures is not unrelated to the central limit theorem of probabilities, although the interesting physics is generally hidden in essential deviations from Gaussian distributions. A few algebraic identities about Gaussian expectation values, in particular Wick’s theorem are recalled. Integrals over some type of formally complex conjugate variables, directly relevant for boson systems are defined. Fermion systems require the introduction of Grassmann or exterior algebras, and the corresponding generalization of the notions of differentiation and integration. Both for complex and Grassmann integrals, Gaussian integrals, and Gaussian expectation values are calculated, and generalized Wick’s theorems proven. The concepts of generating functions and Legendre transformation are recalled.}, booktitle={Quantum Field Theory and Critical Phenomena}, publisher={Oxford University PressOxford}, author={Zinn-Justin, Jean}, year={2021}, month=apr, pages={1–17}}

@book{Altland_Simons_2010, edition={2}, title={Condensed Matter Field Theory}, rights={https://www.cambridge.org/core/terms}, ISBN={9780521769754}, url={https://www.cambridge.org/core/product/identifier/9780511789984/type/book}, DOI={10.1017/CBO9780511789984}, abstractNote={Modern experimental developments in condensed matter and ultracold atom physics present formidable challenges to theorists. This book provides a pedagogical introduction to quantum field theory in many-particle physics, emphasizing the applicability of the formalism to concrete problems. This second edition contains two new chapters developing path integral approaches to classical and quantum nonequilibrium phenomena. Other chapters cover a range of topics, from the introduction of many-body techniques and functional integration, to renormalization group methods, the theory of response functions, and topology. Conceptual aspects and formal methodology are emphasized, but the discussion focuses on practical experimental applications drawn largely from condensed matter physics and neighboring fields. Extended and challenging problems with fully worked solutions provide a bridge between formal manipulations and research-oriented thinking. Aimed at elevating graduate students to a level where they can engage in independent research, this book complements graduate level courses on many-particle theory.}, publisher={Cambridge University Press}, author={Altland, Alexander and Simons, Ben D.}, year={2010}, month=mar }

@misc{zee_quantum_2010,
	title = {Quantum {Field} {Theory} in a {Nutshell} {\textbar} {Princeton} {University} {Press}},
	url = {https://press.princeton.edu/books/hardcover/9780691140346/quantum-field-theory-in-a-nutshell},
	urldate = {2024-05-01},
	month = feb,
	year = {2010},
	note = {ISBN: 9780691140346},
}

@misc{JGRichardson_StefanKrastanov_2025, title={jgr-rgb/genqo: genqo v0.1.1}, rights={MIT License}, url={https://zenodo.org/doi/10.5281/zenodo.17214019}, DOI={10.5281/ZENODO.17214019}, abstractNote={This is the first release of genqo, in parallel with a publication describing the mathematical formalism of the toolbox and it’s integration with QuantumSavory.jl and QuantumSymbolics.jl}, publisher={Zenodo}, author={J. Gabriel Richardson and Stefan Krastanov}, year={2025}, month=sep }

@article{imoto_quantum_1985,
    title = {Quantum nondemolition measurement of the photon number via the optical {Kerr} effect},
    volume = {32},
    copyright = {http://link.aps.org/licenses/aps-default-license},
    issn = {0556-2791},
    url = {https://link.aps.org/doi/10.1103/PhysRevA.32.2287},
    doi = {10.1103/PhysRevA.32.2287},
    language = {en},
    number = {4},
    urldate = {2024-04-15},
    journal = {Physical Review A},
    author = {Imoto, N. and Haus, H. A. and Yamamoto, Y.},
    month = oct,
    year = {1985},
    pages = {2287--2292},
}

\onecolumngrid

\appendix

\newpage
\pagebreak
\newpage
\pagebreak

\section{Mathematical Principles} \label{app:math-details}

In this section, we describe all mathematical details of modeling the source states using a hybrid Gaussian/non-Gaussian formalism. We provide these details in their completeness to support users of the genqo toolbox, and extensions beyond the sources considered here.

To aid navigating this section, here is a guide to its content:
\begin{itemize}
    \item In subsection \ref{sec:gauss} we start with the photonic source. Here we derive the covariance matrices describing the photonic Gaussian states emitted by the sources. Readers familiar with Gaussian quantum state formalism~\cite{weedbrook_gaussian_2012} can skip much of this section.
    \item In subsection \ref{app:k-function} we discuss losses, through a derivation of the density matrix of the photonic state emitted by the source, with loss incorporated, in the coherent state basis. This is achieved using the K-function formalism~\cite{gagatsos_efficient_2019}. Readers familiar with the Husimi-Q function formalism will find this approach familiar. 
    \item In subsection \ref{sec:detection} we present the effects of measurements. This section is a derivation of various non-Gaussian single-photon detection operations on the density operator, including a description of simplification using Wick's theorem. Readers familiar with high-moment-order Gaussian integral calculations will find the techniques in this section familiar. 
    \item In subsection \ref{app:prob-fid} we end the study of the optical domain physics with a derivation of various performance metrics, including the probability of generation and fidelity with respect to the Bell state. 
    \item Finally, in subsection \ref{app:memory} we discuss the storage of these photonic states in matter qubits. We give a description of how Duan-Kimble style quantum memories are incorporated for modeling the spin-spin state. Readers familiar with the Duan-Kimble loading ~\cite{duan_scalable_2004, chen_polarization_2021} will find this section a direct application. 
\end{itemize}

Throughout this section, we also document where each of these concepts is implemented in the genqo software package.

\subsection{Gaussian Model via Covariance Matrices}
\label{sec:gauss}

\textit{All math presented in this section is translated to code in genqo under the method \texttt{calculate\_covariance\_matrix}. This routine returns the covariance matrix for a given source class (e.g.\ \texttt{ZALM} or \texttt{SPDC}), parameterized by the mean photon number.}

Recall that the the creation ($\hat{a}$) and annihilation ($\hat{a}^{\dagger}$) operators act on the Fock and coherent states as
\begin{align}\label{operators}
    \hat{a} \ket{n} &= \sqrt{n} \ket{n-1} \\
    \hat{a}^\dagger \ket{n} &= \sqrt{n+1} \ket{n+1} \\
    \hat{a}^\dagger \hat{a} \ket{n} &= \hat{N} \ket{n} = n \ket{n} \\ 
    \hat{a} \ket{\alpha} &= \alpha \ket{\alpha}
\end{align}
The operators ($\hat{a}$) and ($\hat{a}^{\dagger}$) are related to the quadrature field operators (position $\hat{q}$ and momentum $\hat{p}$) by \citep{weedbrook_gaussian_2012}
\begin{align}
    \hat{q}_k &= \hat{a}_k + \hat{a}_k^{\dagger} \\
    \hat{p}_k &= i(\hat{a}_k - \hat{a}_k^{\dagger}) \\
    \hat{a}_k &= \frac{1}{\sqrt{2}}(\hat{q}_k + i \hat{p}_k)
\end{align}
Where the subscript $k$ labels a specific mode. For working with an N mode system, the vector of the canonical quadrature operators $\mathbf{\hat{x}}$ (which we will call the quad vector) can be defined in what is sometime referred to as the "qqpp" representation
\begin{align}
    \mathbf{\hat{x}} := (\hat{q}_{1},..., \hat{q}_{N}, \hat{p}_{1},...,\hat{p}_{N})^T,
\end{align}
In the "qpqp" representation, the quad vector would be
\begin{align}
    \mathbf{\hat{x}} := (\hat{q}_{1},\hat{p}_{1},\hat{q}_{2},\hat{p}_{2},...,\hat{q}_{N},\hat{p}_{N})^T
\end{align}
The canonical quadrature operators define the covariance matrix $V$,
\begin{align}
    V_{ij} &= \frac{1}{2} \mmnt{ \{ \Delta \hat{x}_i , \Delta \hat{x}_j \} }\\
    \Delta \hat{x}_i &= \hat{x}_i - \mmnt{\hat{x}_i} \\
    \{ \hat{A}, \hat{B} \} &= \hat{A} \hat{B} + \hat{B} \hat{A}
\end{align}
A Gaussian unitary operation on the Gaussian state, represented by a symplectic matrix $S$, is applied to the covariance matrix as
\begin{align}
    V' = S V S^T
\end{align}
For an $n$ mode system, the covariance matrix is $2n \times 2n$. To completely represent a Gaussian unitary operations on any Gaussian quantum state, symplectic operations must also be applied to the first moments, which is called the displacement vector $d = \mmnt{\hat{x}} = \Tr{\hat{x} \rho}$, where $\rho$ is the quantum state of the system, must also be considered. The displacement vector is transformed by
\begin{align}
    d' = S^T d
\end{align}
The states considered in this manuscript experience no displacement operations, or in other words are "zero-mean", so the displacement vector can be ignored in the following formulation. 

\subsubsection{Covariance Matrix of the Two Mode Squeezed Vacuum State} \label{covtmsv}
The nonlinear process of parametric down conversion is mathematically equivalent to the two mode squeezed vacuum (TMSV) state \cite{lvovsky_squeezed_2016}. The covariance matrix of the two mode squeezed vacuum state (in the "qpqp" ordering) is \citep{davis_improved_2022}
\begin{align} \label{vtmsv}
    V_{\text{TMSV}}(\mu) &= \begin{pmatrix}
        \mathbf{A} & \mathbf{B} \\
        \mathbf{B} & \mathbf{A}
    \end{pmatrix} \\
    \mathbf{A} &= \begin{pmatrix}
        1 + 2\mu & 0 \\
        0 & 1 + 2\mu
    \end{pmatrix} \\
    \mathbf{B} &= \begin{pmatrix}
        2 \sqrt{\mu(\mu+1)} & 0 \\
        0 & -2 \sqrt{\mu(\mu+1)}
    \end{pmatrix},
\end{align}
where $\mu$ is the mean photon number per mode of the TMSV state. In the "qqpp" ordering, it is \citep{takeoka_full_2015}
\begin{align}
    V_{\text{TMSV}}(\mu) &= \begin{pmatrix}
        \gamma^+(\mu) & 0 \\
        0 & \gamma^-(\mu)
    \end{pmatrix} \\
    \gamma^{\pm}(\mu) &= \begin{pmatrix}
        2\mu + 1 & \pm 2 \sqrt{\mu(\mu+1)} \\
        \pm 2 \sqrt{\mu(\mu+1)} & 2\mu + 1
    \end{pmatrix}
\end{align}
The following development of the zero-added loss multiplexed source will be in the "qpqp" representation, before converting to the "qqpp" representation for reasons explained later. In any modeling of this kind, keeping track of the representation is a vital detail which cannot be overlooked.

\subsubsection{Covariance Matrix of a Dual-Rail Entanglement source}
The covariance matrix for a common source of dual-rail entanglement \cite{kwiat_new_1995} \cite{kwiat_ultrabright_1999}, which this manuscript will henceforth refer to as the SPDC source, is constructed by 
\begin{itemize}
    \item Considering two, two-mode squeezed vacuum states. For covariance matrices, this is accomplished by taking the direct sum of two, TMSV covariance matrices, as developed in the previous section \ref{covtmsv}.
    \item Performing a mode swap of the idler modes. For covariance matrices, this is accomplished by applying permutation matrices.
\end{itemize}
The quad vector for a single TMSV covariance matrix in the "qpqp" representation is 
\begin{align}
    \vec{x} = \{q_1, p_1, q_2, p_2\}
\end{align}
Labeling the first TMSV as A and the second as B, the quad vector for the covariance matrix of two, TMSV states is
\begin{align}
    \vec{x} = \{q_{1A}, p_{1A}, q_{2A}, p_{2A}, q_{1B}, p_{1B}, q_{2B}, p_{2B}\}
\end{align}
Mode swapping will occur between idler modes (i.e. mode 2 of each TMSV state), therefore the columns and rows of the covariance matrix will be swapped such that
\begin{align*}
    q_{2A} &\Leftrightarrow q_{2B} \\
    p_{2A} &\Leftrightarrow p_{2B}
\end{align*}
The permutation matrix used to do the mode swapping is
\begin{align} \label{vspdc}
    V_{spdc} &= P (V_{tmsv} \oplus V_{tmsv}) P^{T} \\
    P &= \begin{bmatrix}
        1 & 0 & 0 & 0 & 0 & 0 & 0 & 0 \\
        0 & 1 & 0 & 0 & 0 & 0 & 0 & 0 \\
        0 & 0 & 0 & 0 & 0 & 0 & 1 & 0 \\
        0 & 0 & 0 & 0 & 0 & 0 & 0 & 1 \\
        0 & 0 & 0 & 0 & 1 & 0 & 0 & 0 \\
        0 & 0 & 0 & 0 & 0 & 1 & 0 & 0 \\
        0 & 0 & 1 & 0 & 0 & 0 & 0 & 0 \\
        0 & 0 & 0 & 1 & 0 & 0 & 0 & 0
    \end{bmatrix}
\end{align}
The result is the complete covariance matrix of the SPDC source, with the final quad vector in the "qpqp" representation
\begin{align}
    \vec{x} = (q_1, p_1, q_2, p_2, q_3, p_3, q_4, p_4)
\end{align}

\subsubsection{Initial Covariance Matrix of the Cascaded/ZALM Source}
The single-mode cascaded source and the single-mode ZALM source are equivalent, and hence the formulation of the covariance matrices in this section apply for both cases. There exist some nuances for modeling the frequency multiplexing components of the ZALM source which are not explored in this paper.

The initial covariance matrix of the cascaded source $V_{casc,i}$ is related to the covariance matrix of the SPDC source $V_{spdc}$ by 
\begin{align}
    V_{casc,i} = P_3 (V_{spdc} \oplus V_{spdc}) P_3^T
\end{align}
Where $V_{spdc}$ follows from \ref{vspdc} and $P_3$ is a permutation matrix that ensures the covariance matrix is in "qqpp" representation. This translation from "qpqp" to "qqpp" is done for two reasons. First, the expressions for the K-function formalism as presented in \cite{gagatsos_efficient_2019} are in "qqpp". Second, the heralding Bell state measurement which will be performed next is slightly less complicated when considered in "qqpp".

Central to the functionality of the cascaded/ZALM source is a Bell state measurement which heralds the creation of dual-rail Bell states. Beam splitters, required for the Bell state measurement between the last two modes of the first SPDC source with the first two modes of the second SPDC source, can be applied via symplectic matrices. 

\subsubsection{Bell State Measurement}
To understand how these beam splitter operations will occur, consider for example the SPDC sources in the sagnac configuration. This means that the 8 modes of $V_{casc,i}$ are spatio-polarization modes, with each pair of modes being the same spatial mode, and within each pair existing a horizontal and a vertical polarization mode. The basis state can be rewritten as
$$
\ket{H_{1a},V_{1a},H_{2a},V_{2a},H_{1b},V_{1b}H_{2b},V_{2b}}
$$
Where subscripts $a$ and $b$ are the SPDC source label, and $1$ and $2$ label the spatial mode. Therefore, the beam splitters for performing Bell state measurements will occur between modes $\{H_{2a},H_{1b}\}$ and modes $\{V_{2a},V_{1b}\}$. In correspondence to the previously considered mode basis, this means beampslitters will occur between modes $\{3,5\}$ and modes $\{4,6\}$

Beam splitters play a key role in both the source and the Duan-Kimble memory encoding processes. A beamsplitter unitary between modes a and b \citep{weedbrook_gaussian_2012}
\begin{align}
    U_{bs} = e^{\theta(\hat{a}^{\dagger} \hat{b} - \hat{a} \hat{b}^{\dagger})}
\end{align}
Where the transmissivity $\tau = \cos^2(\theta)$.

In the symplectic matrix form, beam splitters on modes A and B with transmissivity $t$ is \citep{takeoka_full_2015}
\begin{align}
    S_{AB}^{(t)} = \begin{pmatrix}
        \sqrt{t} & \sqrt{1 - t} & 0 & 0 \\
        -\sqrt{1 - t} & \sqrt{t} & 0 & 0 \\
        0 & 0 & \sqrt{t} & \sqrt{1 - t} \\
        0 & 0 & -\sqrt{1 - t} & \sqrt{t} \\
    \end{pmatrix}
\end{align}
For a multi mode system of N total modes, a beam splitter between modes $n$ and $m$ would be a product sum $\bigoplus_{i = 1}^N$, where every element is the identity operator other than the $n$-th and $m$-th elements would be $S^{(t)}$, where 
\begin{align} \label{symplec}
    S^{(t)} = \begin{pmatrix}
        \sqrt{t} & \sqrt{1 - t}  \\
        -\sqrt{1 - t} & \sqrt{t} \\
    \end{pmatrix}
\end{align}
The symplectic form of beamsplitters can be valuable in the context of the ZALM source, where interference is required for Bell state measurements, and representation can be simplified when this is performed on the Gaussian state rather than on the K-function represented density matrix. Duan-Kimble memory loading will require polarizing beam splitter operations on pairs of adjacent coherent state represented modes. 

Represented as symplectic matrices, this means
\begin{align}
    V_{casc} = S_{4,6} S_{3,5} V_{casc,i} S_{3,5}^T S_{4,6}^T
\end{align}
Where $S_{i,j}$ represents a symplectic matrix that performs a 50/50 beamsplitter operation between modes $i$ and $j$. Because the quad vector for the covariance matrix is $\ket{q_{1}, q_{2}, q_{3}, q_{4}, q_{5}, q_{6}, q_{7}, q_{8}, p_{1}, p_{2}, p_{3}, p_{4}, p_{5}, p_{6}, p_{7}, p_{8}}$
\begin{align}
    S_{3,5} &= \mathbf{I}_{2} \oplus \begin{bmatrix}
        \sqrt{t} & 0 & \sqrt{1 - t} & 0 \\
        0 & 1 & 0 & 0 \\
        -\sqrt{1 - t} & 0 & \sqrt{t} & 0 \\
        0 & 0 & 0 & 1 \\
    \end{bmatrix} \oplus \mathbf{I}_{2} \oplus \mathbf{I}_{2} \oplus \begin{bmatrix}
        \sqrt{t} & 0 & \sqrt{1 - t} & 0 \\
        0 & 1 & 0 & 0 \\
        -\sqrt{1 - t} & 0 & \sqrt{t} & 0 \\
        0 & 0 & 0 & 1 \\
    \end{bmatrix} \oplus \mathbf{I}_{2} \\
    S_{4,6} &= \mathbf{I}_{2} \oplus \begin{bmatrix}
        1 & 0 & 0 & 0 \\
        0 & \sqrt{t} & 0 & \sqrt{1 - t} \\
        0 & 0 & 1 & 0 \\
        0 & -\sqrt{1 - t} & 0 & \sqrt{t} \\
    \end{bmatrix} \oplus \mathbf{I}_{2} \oplus \mathbf{I}_{2} \oplus \begin{bmatrix}
        1 & 0 & 0 & 0 \\
        0 & \sqrt{t} & 0 & \sqrt{1 - t} \\
        0 & 0 & 1 & 0 \\
        0 & -\sqrt{1 - t} & 0 & \sqrt{t} \\
    \end{bmatrix} \oplus \mathbf{I}_{2}
\end{align}

\onecolumngrid
 
\subsection{From Gaussian to Density Matrix}
\label{app:k-function}
In this section we describe the process of transforming a state represented as a covariance matrix into a density matrix representation via the K-function formalism. 

\subsubsection{$K$-Function Formalism} \label{kfun}

\textit{ These concepts are implemented in genqo through \texttt{calculate\_k\_function\_matrix} which takes the covariance matrix and provides the K-function matrix, an intermediary step to the computation of the density matrix.}

The K-function formalism \cite{gagatsos_efficient_2019} translates a covariance matrix into a density matrix in the coherent state basis. This translation allows non-Gaussian operations to be applied to the state. 

Consider $N$-bosonic modes with canonical position and momentum operators $\hat{q}_k,\hat{p}_k;\; k\in\{1,2,\ldots,N\}$. Define $N$-mode coherent states as $\ket{\vec{\alpha}} = \ket{\alpha_1,\alpha_2,\ldots,\alpha_N}; \; \alpha_k\in \mathbb{C}$. Each coherent amplitude can be expressed in terms of the canonical quadrature operators $\alpha_k= \frac{1}{\sqrt{2}}(q_k+i p_k)$, i.e.\ the real and imaginary parts of $\alpha_k$ correspond to displacements along the position and momentum quadratures of the $k$-th mode. An $N$-mode pure quantum state $\ket{\psi}$ in $K$-function representation (i.e.\ projection on the coherent state resolution of the identity operator) is
\begin{align} 
    \begin{split}
        \ket{\psi} &= \frac{1}{(2\pi)^N} \int d^{2N} \vec{x}_{\alpha} \ket{\vec{\alpha}}\!\!\braket{\vec{\alpha}|\psi} =\int d^{2N} \vec{x}_{\alpha} K(\vec{x}_{\alpha}) \ket{\vec{\alpha}} 
    \end{split}
\end{align}
where $K(\vec{x}_{\alpha}) = (2\pi)^{-N}\braket{\vec{\alpha}|\psi} $. If $\ket{\psi}$ is a zero mean $N$-mode Gaussian state (i.e. the state is completely specified by a covariance matrix $V$), then $ K(\vec{x_{\alpha}})$ can be expressed as,
\begin{align}
    K(\vec{x}_{\alpha}) &= \frac{\exp{\left[-\frac{1}{2} \vec{x}_{\alpha}^T \mathcal{B} \vec{x}_{\alpha} \right]}}{(2\pi)^N (\det{\Gamma})^{1/4}},
\end{align}
where,
\begin{subequations}
    \begin{align}
    \Gamma &= V + I/2 \\
    \mathcal{B} &= \frac{1}{2} \begin{pmatrix}
        A + \frac{i}{2}(C + C^T) & C - \frac{i}{2}(A-B) \\
        C^T - \frac{i}{2}(A - B) & B - \frac{i}{2}(C+C^T)
    \end{pmatrix} \\
    \vec{x}_{\alpha} &= (\vec{q}_{\alpha},\vec{p}_{\alpha}) = (q_1,q_2,\ldots,q_N,p_1,p_2,\ldots,p_N)\\
    \Gamma^{-1} &= \begin{pmatrix}
        A & C \\
        C^T & B
    \end{pmatrix}    
\end{align}
\end{subequations}

\subsubsection{Incorporating Loss} \label{loss}
Loss can be incorporated into the quantum state in the K-function formalism using Kraus operators ($A_k$) \citep{dhara_heralded_2022}, where
\begin{align}
    \hat{A_k} &= \sqrt{\frac{(1-\eta)^k}{k!}} \sqrt{\eta}^{\hat{n}} \hat{a}^k \\
    \hat{A_k}^{\dagger} &= \sqrt{\frac{(1-\eta)^k}{k!}} (\hat{a}^k)^{\dagger} \sqrt{\eta}^{\hat{n}}
\end{align}
These acts on coherent basis terms $\ket{\gamma}\!\!\bra{\delta}$, where $\gamma,\delta \in \mathbb{C}$ as
\begin{align}
    &\sum_{k = 0}^{\infty} \hat{A_k} \ket{\gamma}\!\!\bra{\delta} \hat{A_k}^{\dagger} \nonumber \\
    &= \sum \frac{(1 - \eta)^k}{k!} \sqrt{\eta}^{\hat{n}} \hat{a}^k  \ket{\gamma}\!\!\bra{\delta} \hat{a}^{\dagger k} \sqrt{\eta}^{\hat{n}}  \nonumber \\
    &= \sum \frac{(1 - \eta)^k}{k!} (\gamma \delta^*)^k \nonumber \\
    &\exp{\left( -\frac{(|\gamma|^2) + |\delta|^2))(1 - \eta)}{2} \right)} \ket{\gamma \sqrt{\eta}} \bra{\delta \sqrt{\eta}} \nonumber \\
    &= \exp{\left( (\gamma \delta^*)(1 - \eta) - \frac{(|\gamma|^2) + |\delta|^2))(1 - \eta)}{2} \right)} \nonumber \\
    &\ket{\gamma \sqrt{\eta}} \bra{\delta \sqrt{\eta}} \\
    &= \mathcal{G}(\gamma, \delta) \ket{\gamma \sqrt{\eta}} \bra{\delta \sqrt{\eta}}
\end{align}
Recalling that $\exp{(x)} = \sum_{k = 0}^{\infty}{x^k}/{k!}$.

\subsubsection{Single-mode ZALM Source State}
The previously developed cascaded source state is translated into the coherent state basis using the K-function formalism. Kraus operators incorporate channel loss ($\eta_t$), detector loss ($\eta_d$), and Bell measurement loss ($\eta_b$)  onto the coherent basis states, as described in \cite{dhara_heralded_2022}, resulting in the density matrix
\begin{align} \label{zstate}
    \rho_{zl} = \int d^{16} \vec{x_{\alpha}} d^{16} \vec{x_{\beta}} K(\vec{x_{\alpha}}) K(\vec{x_{\beta}})^* \mathcal{G}(\Vec{\alpha}, \Vec{\beta}, \Vec{\eta}) \ket{\Vec{\alpha}_L} \bra{\Vec{\beta}_L}  
\end{align}
Where
\begin{align}
    \Vec{\alpha} &= (\alpha_1; \alpha_2; \alpha_3; \alpha_4; \alpha_5; \alpha_6; \alpha_7; \alpha_8)  \\
    \Vec{\beta} &= (\beta_1; \beta_2; \beta_3; \beta_4; \beta_5; \beta_6; \beta_7; \beta_8) \\
    \Vec{\eta} &= (\eta_t \eta_d; \eta_t \eta_d; \eta_b; \eta_b; \eta_b; \eta_b; \eta_t \eta_d; \eta_t \eta_d;) \\
    \Vec{\alpha}_L &= \begin{cases}
        \alpha_i \sqrt{\eta_i} \text{ for } i = 1,2,7,8 \\
        \alpha_i \sqrt{\eta_i} \text{ for } i = 3,4,5,6
    \end{cases} \\
    \Vec{\beta}_L &= \begin{cases}
        \beta_i \sqrt{\eta_i} \text{ for } i = 1,2,7,8 \\
        \beta_i \sqrt{\eta_i} \text{ for } i = 3,4,5,6
    \end{cases}\\
    \mathcal{G}(\Vec{\alpha}, \Vec{\beta}, \Vec{\eta}) &= \prod_{i=1}^8 \exp((\alpha_i \beta^*_i)(1 - \eta_i) - \frac{(|\alpha_i|^2 + |\beta_i|^2)(1 - \eta_i)}{2}) 
\end{align}
The following sections will develop non-Gaussian detection onto this state, and simplification to calculate metrics which are useful for analyzing the entanglement distribution performance of the state under certain conditions.

\onecolumngrid

\subsection{Detection}
\label{sec:detection}
Consider an arbitrary detection projector on the 8 modes of a single cascaded source photonic state. 
\begin{align}
    \Pi_d = \bigotimes_{i=1}^8 \outprod{d_i}
\end{align}
Where $\ket{d_i}$ is a fock state. Considering the action of this projection operator a density matrix in the coherent state basis will be important for future analysis. Recalling \ref{zstate}
\begin{align}
    \Tr( \Pi_d \rho_{zl}) &= \int d^{16} \vec{x_{\alpha}} d^{16} \vec{x_{\beta}} K(\vec{x_{\alpha}}) K(\vec{x_{\beta}})^* \mathcal{G}(\Vec{\alpha}, \Vec{\beta}, \Vec{\eta}) \left( \prod_{j=1}^8 \ket{d_j} \inprod{d_j}{\alpha_j \sqrt{\eta_j}} \bra{\beta_j \sqrt{\eta_j} }  \right) \nonumber \\
    &= \sum_{l = 0}^{\infty} \bra{l} \left[ \int d^{16} \vec{x_{\alpha}} d^{16} \vec{x_{\beta}} K(\vec{x_{\alpha}}) K(\vec{x_{\beta}})^* \mathcal{G}(\Vec{\alpha}, \Vec{\beta}, \Vec{\eta}) \left( \prod_{j=1}^8 \ket{d_j} \inprod{d_j}{\alpha_j \sqrt{\eta_j}} \bra{\beta_j \sqrt{\eta_j} }  \right) \right] \ket{l} \nonumber \\
    &= \int d^{16} \vec{x_{\alpha}} d^{16} \vec{x_{\beta}} K(\vec{x_{\alpha}}) K(\vec{x_{\beta}})^* \mathcal{G}(\Vec{\alpha}, \Vec{\beta}, \Vec{\eta}) \left( \prod_{j=1}^8  \inprod{d_j}{\alpha_j \sqrt{\eta_j}} \inprod{\beta_j \sqrt{\eta_j} }{d_j}  \right)
\end{align}
Where the inner products simplify as
\begin{align}
    \inprod{d_i}{\alpha_j \sqrt{\eta_j} }&= \nonumber \\
    & e^{-|\alpha_j \sqrt{\eta_j}|^2/2} \frac{(\alpha_j \sqrt{\eta_j})^{d_i}}{\sqrt{d_i!}} \\
    \inprod{d_i}{\alpha_j \sqrt{\eta_j} }&\inprod{\beta_j \sqrt{\eta_j}}{d_i}= \nonumber \\
    & e^{-\frac{1}{2}\left( |\alpha_j \sqrt{\eta_j}|^2 + |\beta_j \sqrt{\eta_j}|^2 \right)} \frac{(\alpha_j \beta_j^* \eta_j)^{d_i}}{d_i!}
\end{align}
So, the trace expression simplifies to
\begin{align}
    Tr(\rho_{z} \Pi_d) &= \int d^{16} \vec{x_{\alpha}} d^{16} \vec{x_{\beta}} K(\vec{x_{\alpha}}) K(\vec{x_{\beta}})^* \mathcal{G}(\Vec{\alpha}, \Vec{\beta}, \Vec{\eta}) \nonumber \\
    & \times e^{-\frac{1}{2}\left( \sum_{j=1}^{8} |\alpha_j \sqrt{\eta_j}|^2 + |\beta_j \sqrt{\eta_j}|^2 \right)} \left( \prod_{j=1}^8  \frac{(\alpha_j \beta_j^* \eta_j)^{d_i}}{d_i!}  \right) 
\end{align}
Observe that no matter the detection patter, there will always exist a factor of $e^{-\frac{1}{2}\left( \sum_{j = 1}^{8}|\alpha_j \sqrt{\eta_j}|^2 + |\beta_j \sqrt{\eta_j}|^2 \right)}$. Together with the K functions and the loss $\mathcal{G}$ function, this factor will constitute what will be referred to as the "$\mathbf{A}$" matrix, which we will describe in the next section. For simplification, define the coefficient
\begin{align}
    C(\vec{d}) = \left( \prod_{j=1}^8  \frac{(\alpha_j \beta_j^* \eta_j)^{d_i}}{d_i!}  \right) 
\end{align}

\subsubsection{Fock Basis Density Matrix Elements}
Projection onto the Fock basis can enable us to calculate the complete density matrix. Consider an aribitrary projection 
\begin{align} 
    \bra{\vec{d}} \rho_{zl} \ket{\vec{g}} &= \int d^{16} \vec{x_{\alpha}} d^{16} \vec{x_{\beta}} K(\vec{x_{\alpha}}) K(\vec{x_{\beta}})^* \mathcal{G}(\Vec{\alpha}, \Vec{\beta}, \Vec{\eta}) \left( \prod_{j=1}^8  \inprod{d_j}{\alpha_j \sqrt{\eta_j}} \inprod{\beta_j \sqrt{\eta_j} }{g_j}  \right) \\
    &= \int d^{16} \vec{x_{\alpha}} d^{16} \vec{x_{\beta}} K(\vec{x_{\alpha}}) K(\vec{x_{\beta}})^* \mathcal{G}(\Vec{\alpha}, \Vec{\beta}, \Vec{\eta}) e^{-\frac{1}{2}\left( \sum_{j=1}^{8} |\alpha_j \sqrt{\eta_j}|^2 + |\beta_j \sqrt{\eta_j}|^2 \right)} \left( \prod_{j=1}^8  \frac{(\alpha_j \sqrt{\eta_j})^{d_j}}{\sqrt{d_j!}}\frac{(\beta_j \sqrt{\eta_j})^{g_j}}{\sqrt{g_j!}}  \right)
\end{align}
Therefore, the $\mathbf{A}$ matrix remains unchanged, but the $C$ coefficient becomes
\begin{align}
    C(\vec{d},\vec{g}) = \left( \prod_{j=1}^8  \frac{(\alpha_j \sqrt{\eta_j})^{d_j}}{\sqrt{d_j!}}\frac{(\beta_j \sqrt{\eta_j})^{g_j}}{\sqrt{g_j!}}  \right)
\end{align}

\subsubsection{Calculating the $\mathbf{A}$ matrix} \label{ZALMamat}
For a single cascaded source, the components that make up the $\mathbf{A}$ matrix are
\begin{align}
    K_z(\vec{x_{\alpha}}) K_z(\vec{x_{\beta}})^* &\mathcal{G}(\Vec{\alpha}, \Vec{\beta}, \Vec{\eta}) e^{-\frac{1}{2}\left( \sum_{j = 1}^{8}\eta_j (|\alpha_j|^2 + |\beta_j|^2) \right)}
\end{align}
Recall that
\begin{align}
    \Vec{\eta} &= (\eta_1 ; \eta_2; \eta_3; \eta_4; \eta_5; \eta_6; \eta_7; \eta_8) \nonumber \\
    &= (\eta_t \eta_d; \eta_t \eta_d; \eta_b; \eta_b; \eta_b; \eta_b; \eta_t \eta_d; \eta_t \eta_d;) \\
    \mathcal{G}(\Vec{\alpha}, \Vec{\beta}, \Vec{\eta}) &= \prod_{i=1}^8 \exp((\alpha_i \beta^*_i)(1 - \eta_i) - \frac{(|\alpha_i|^2 + |\beta_i|^2)(1 - \eta_i)}{2})
\end{align}
Therefore, the exponential components that constitute the $\mathbf{A}$ matrix simplify to 
\begin{align} \label{amat}
    &K_z(\vec{x_{\alpha}}) K_z(\vec{x_{\beta}})^* \exp\left[\sum_{i=1}^8(\alpha_i \beta^*_i)(1 - \eta_i) - \frac{(|\alpha_i|^2 + |\beta_i|^2)(1 - \eta_i)}{2}\right] \exp\left[-\frac{1}{2}\left( \sum_{j = 1}^{8}\eta_j(|\alpha_j|^2 + |\beta_j|^2) \right) \right] \\ \nonumber
    &= K_z(\vec{x_{\alpha}}) K_z(\vec{x_{\beta}})^* \exp\left[-\frac{1}{2} \left( \sum_{i=1}^8 (2\eta_i - 2)(\alpha_i \beta^*_i) + (1 - \eta_i)(|\alpha_i|^2 + |\beta_i|^2)\right) \right] \exp\left[ -\frac{1}{2}\left( \sum_{j = 1}^{8}(\eta_j)(|\alpha_j|^2 + |\beta_j|^2) \right) \right] \\ \nonumber
    &= K_z(\vec{x_{\alpha}}) K_z(\vec{x_{\beta}})^* \exp \left[ -\frac{1}{2}\left( \sum_{j = 1}^{8} (2\eta_i - 2)(\alpha_i \beta^*_i) + (|\alpha_i|^2 + |\beta_i|^2) \right) \right] \\ \nonumber 
    &\propto  \exp\left[ -\frac{1}{2}\vec{x}(\mathcal{B}_{\alpha} \oplus \mathcal{B}^*_{\beta})\vec{x}^T \right] \exp\left[ -\frac{1}{2}\left( \sum_{j = 1}^{8} (2\eta_i - 2)(\alpha_i \beta^*_i) + (|\alpha_i|^2 + |\beta_i|^2) \right)\right] \\ \nonumber 
    &= \exp\left[-\frac{1}{2}\vec{x}(\mathcal{B}_{\alpha} \oplus \mathcal{B}^*_{\beta})\vec{x}^T \right] \exp \left[ -\frac{1}{2}\left( \sum_{i = 1}^{8} (\eta_i - 1)(q_{\alpha_i} + i p_{\alpha_i})(q_{\beta_i} - i p_{\beta_i}) + \frac{1}{2}(|q_{\alpha_i}|^2 +  |p_{\alpha_i}|^2 + |q_{\beta_i}|^2 +  |p_{\beta_i}|^2) \right) \right] \\ \nonumber 
    &= \exp \left[ -\frac{1}{2} \vec{x} \mathbf{A} \vec{x}^T \right]
\end{align}
Where
\begin{align}
    \vec{x} &= \vec{x}_{\alpha} \bigcup \vec{x}_{\beta} \nonumber \\
    &= \{ q_{1,\alpha}, ..., q_{8,\alpha}, p_{1,\alpha}, ..., p_{8,\alpha}, q_{1,\beta}, ..., q_{8,\beta}, p_{1,\beta}, ..., p_{8,\beta}  \}
\end{align}
Therefore, for any detection pattern $\Pi_d$, 
\begin{align}
    &\Tr(\rho_{z} \Pi_d) = \nonumber \\
    &\frac{1}{(2\pi)^{16} (\det{\Gamma})^{1/4} (\det{\Gamma^*})^{1/4}} \int d^{32} \vec{x} C(\vec{d}) \exp \left[ -\frac{1}{2} \vec{x} \mathbf{A} \vec{x}^T \right]
\end{align}
A portion of this expression can be further simplified using Wick's Theorem as
\begin{align}
    \frac{1}{(2 \pi )^{16}}\int d^{32} \mathbf{x} C e^{-\frac{1}{2} \mathbf{x}^T \mathbf{A} \mathbf{x}} = \frac{1}{\sqrt{\det{\mathbf{A}}}} \mathbf{W}(\mathbf{A},C,\vec{x})
\end{align}
Which is fully derived later in section \ref{app:wick}. In this, we get to the fact that the $\mathbf{A}$ matrix is a central argument to performing the Gaussian integration via Wick's theorem.

\onecolumngrid

\subsubsection{Derivation of Wick's Theorem}
\label{app:wick}
Consider the complex gaussian integral, which states that
\begin{align}
    \int d(\Bar{z},z) e^{- \Bar{z} w z} = \frac{\pi}{w}
\end{align}
Where $z = x + iy$, $\Bar{z}$ is the complex conjugate of of $z$, $\Re(w) > 0$ and $\int d(\Bar{z},z) \equiv \int_{-\infty}^{\infty} dx 
\, dy$. This can be generalized to more than one dimension, by defining an $N$-component vector \begin{align}
    \int d(\mathbf{v}^{\dagger}, \mathbf{v}) e^{-\mathbf{v}^{\dagger} \mathbf{A} \mathbf{v}},
\end{align}
where $\mathbf{A}$ is a complex Hermitian matrix. To solve this integral, consider that $\mathbf{A}$ can be diagonalized as $\mathbf{A} = \mathbf{O}^T \mathbf{D} \mathbf{O}$. We can now define $\mathbf{u} = \mathbf{O} \mathbf{v}$ so that the integral becomes
\begin{align}
    \int d(\mathbf{u}^{\dagger}, \mathbf{u}) e^{-\mathbf{u}^{\dagger} \mathbf{D} \mathbf{u}}
\end{align}
Because $D$ is diagonal and $\det(\mathbf{O}) = 1$ this simplifies into a factor of complex gaussian integrals, such that
\begin{align}
    \int d(\mathbf{v}^{\dagger}, \mathbf{v}) e^{-\mathbf{v}^{\dagger} \mathbf{A} \mathbf{v}} = \frac{\pi^N}{\prod_{i=1}^N d_i} = \frac{\pi^N}{\det(\mathbf{D})}= \frac{\pi^N}{\det(\mathbf{A})}
\end{align}
This can be further generalized, such that
\begin{align}
    \int d(\mathbf{v}^{\dagger}, \mathbf{v}) e^{-\mathbf{v}^{\dagger} \mathbf{A} \mathbf{v} + \mathbf{w}^{\dagger} \mathbf{v} + \mathbf{v}^{\dagger} \mathbf{w}'} = \frac{\pi^N}{\det(\mathbf{A})} e^{\mathbf{w}^{\dagger} \mathbf{A}^{-1} \mathbf{w}'}
\end{align}
Where $\mathbf{A}^{-1}$ is the inverse of the A matrix and $\mathbf{w}$ and $\mathbf{w}'$ are independent complex vectors. This identity is resolved such that making the change of variables
\begin{align}
    \mathbf{v} &= \mathbf{u} + \mathbf{A}^{-1} \mathbf{w}' \\
    \mathbf{v}^{\dagger} &= \mathbf{u}^{\dagger} + \mathbf{w}^{\dagger}(\mathbf{A}^{-1})^\dagger
\end{align}
If $\mathbf{A}^{-1}$ is hermitian, then $(\mathbf{A}^{-1})^\dagger = \mathbf{A}^{-1}$, therefore
\begin{align}
    -\mathbf{v}^{\dagger} \mathbf{A} \mathbf{v} + \mathbf{w}^{\dagger} \mathbf{v} + \mathbf{v}^{\dagger} \mathbf{w}' 
    &= -(\mathbf{u}^{\dagger} + \mathbf{w}^{\dagger}(\mathbf{A}^{-1})^\dagger) \mathbf{A} (\mathbf{u} + \mathbf{A}^{-1} \mathbf{w}') + \mathbf{w}^{\dagger} (\mathbf{u} + \mathbf{A}^{-1} \mathbf{w}') + (\mathbf{u}^{\dagger} + \mathbf{w}^{\dagger}(\mathbf{A}^{-1})^\dagger)\mathbf{w}'\\
    &= -\mathbf{u}^{\dagger} \mathbf{A} \mathbf{u} -\mathbf{u}^{\dagger} \mathbf{A} \mathbf{A}^{-1} \mathbf{w}' - \mathbf{w}^{\dagger}(\mathbf{A}^{-1})^\dagger \mathbf{A} \mathbf{u} - \mathbf{w}^{\dagger}(\mathbf{A}^{-1})^\dagger \mathbf{A} \mathbf{A}^{-1} \mathbf{w}' \nonumber \\
    &+ \mathbf{w}^{\dagger} \mathbf{u} + \mathbf{w}^{\dagger} \mathbf{A}^{-1} \mathbf{w}' + \mathbf{u}^{\dagger} \mathbf{w}' + \mathbf{w}^{\dagger}(\mathbf{A}^{-1})^\dagger \mathbf{w}'\\
    &= -\mathbf{u}^{\dagger} \mathbf{A} \mathbf{u} -\mathbf{u}^{\dagger} \mathbf{w}' - \mathbf{w}^{\dagger} \mathbf{u} - \mathbf{w}^{\dagger}(\mathbf{A}^{-1})^\dagger \mathbf{w}' \nonumber + \mathbf{w}^{\dagger} \mathbf{u} + \mathbf{w}^{\dagger} \mathbf{A}^{-1} \mathbf{w}' + \mathbf{u}^{\dagger} \mathbf{w}' + \mathbf{w}^{\dagger}(\mathbf{A}^{-1})^\dagger \mathbf{w}'\\
    &= -\mathbf{u}^{\dagger} \mathbf{A} \mathbf{u} + \mathbf{w}^{\dagger} \mathbf{A}^{-1} \mathbf{w}'
\end{align}
Therefore, 
\begin{align}
    &\int d(\mathbf{v}^{\dagger}, \mathbf{v}) e^{-\mathbf{v}^{\dagger} \mathbf{A} \mathbf{v} + \mathbf{w}^{\dagger} \mathbf{v} + \mathbf{v}^{\dagger} \mathbf{w}'}  \nonumber \\
    &= \int d(\mathbf{u}^{\dagger}, \mathbf{u}) e^{-\mathbf{u}^{\dagger} \mathbf{A} \mathbf{u} + \mathbf{w}'^{\dagger} \mathbf{A}^{-1} \mathbf{w}'} \\
    &= \int d(\mathbf{u}^{\dagger}, \mathbf{u}) e^{-\mathbf{u}^{\dagger} \mathbf{A} \mathbf{u}} e^{\mathbf{w}'^{\dagger} \mathbf{A}^{-1} \mathbf{w}'} \\
    &= \frac{\pi^N}{\det(\mathbf{A})} e^{ \mathbf{w}^{\dagger} \mathbf{A}^{-1} \mathbf{w}'}
\end{align}
The proof of this can be found here~\cite{Palkanoglou_2018}. Consider that
\begin{align}
    &\left. \frac{\partial}{\partial w_m} \frac{\partial}{\partial w'_n}  \int d(\mathbf{v}^{\dagger}, \mathbf{v}) e^{-\mathbf{v}^{\dagger} \mathbf{A} \mathbf{v} + \mathbf{w}^{\dagger} \mathbf{v} + \mathbf{v}^{\dagger} \mathbf{w}'} \right|_{\mathbf{w}=\mathbf{w}' = 0} \nonumber \\
    &= \int d(\mathbf{v}^{\dagger}, \mathbf{v}) e^{-\mathbf{v}^{\dagger} \mathbf{A} \mathbf{v}} v_m \Bar{v_n} \\
    &= \left. \frac{\pi^N}{\det(\mathbf{A})} e^{\mathbf{w}^{\dagger} \mathbf{A}^{-1} \mathbf{w}'} \right|_{\mathbf{w} = \mathbf{w}' = 0} \\
    &= \frac{\pi^N}{\det(\mathbf{A})} A^{-1}_{mn}
\end{align}
These results are connected to the Gaussian moment, and explicitly
\begin{align}
    \langle v_m \Bar{v_n} \rangle = A^{-1}_{mn} = \frac{\det(\mathbf{A})}{\pi^N} \int d(\mathbf{v}^{\dagger}, \mathbf{v}) e^{-\mathbf{v}^{\dagger} \mathbf{A} \mathbf{v}} v_m \Bar{v_n}
\end{align}
Taking more partial derivatives we can calculate higher order moments, and by the chain rule we will get that 
\begin{align}
    \langle \Bar{v_{i_1}} \Bar{v_{i_2}} ... \Bar{v_{i_n}} v_{j_1} v_{j_2} ... v_{j_n} \rangle = \sum_{P} A^{-1}_{j_1, i_{P1}} ... A^{-1}_{j_n, i_{Pn}}
\end{align}
Where $\sum_P$ represents the sum over all permutations of $n$ integers. This property is called Wick Coupling, and this formalism for calculating higher-order Gaussian moments is generally called Wick's theorem, a technique most commonly used in quantum field theory ~\cite{zee_quantum_2010, Zinn-Justin_2021,Altland_Simons_2010}. 

Consider how things change slightly in the real case. Consider the Gaussian integral
\begin{align}
    \int_{-\infty}^{\infty} dx e^{-\frac{a}{2}x^2} = \sqrt{\frac{2 \pi}{ a }}
\end{align}
Where $\Re[{a}] > 0$. This can be extended to multiple dimensions, considering $\int d \mathbf{v} e^{-\frac{1}{2} \mathbf{v}^T \mathbf{A} \mathbf{v}}$ where $\mathbf{A}$ is a positive definite, real, symmetric, $N$-dimensional matrix, and $\mathbf{v}$ is an $N$-component real vector. Because $\mathbf{A}$ is symmetric, it can be diagonalized by orthogonal transformation, $\mathbf{A} = \mathbf{O}^T \mathbf{D} \mathbf{O}$ where the $\mathbf{O}$ matrix is orthogonal and all elements of the diagonal matrix $\mathbf{D}$ are positive. Absorbing $\mathbf{O}$ in to the vector $\mathbf{v}$ the Gaussian integral becomes diagonal, and is just a sequence of Gaussian integrals, therefore
\begin{align}
    \int d \mathbf{v} e^{-\frac{1}{2} \mathbf{v}^T \mathbf{A} \mathbf{v}} = \prod_{i = 1}^N \sqrt{\frac{2 \pi}{d_i}} = \sqrt{\frac{(2 \pi)^N}{\det{\mathbf{D}}}} = \sqrt{\frac{(2 \pi)^N}{\det{\mathbf{A}}}}
\end{align}
The Wick coupling will also change slightly. Following the same procedure as in the complex case, we find that
\begin{align}
    \langle x_1 ... x_n \rangle = \sum_{\text{all possible pairings P of } \{1,...,n\} } \langle x_{P_1} x_{P_2} \rangle ... \langle x_{P_{n-1}} x_{P_n} \rangle
\end{align}

\subsubsection{Application to hybrid Gaussian/Non-Gaussian modeling}
For the purposes of doing hybrid Gaussian/Non-Gaussian modeling, we will need to solve integrals of the form 
\begin{align} \label{genint}
    \int d^{4N} \vec{\mathbf{x}} C e^{-\frac{1}{2} \vec{\mathbf{x}}^T \mathbf{A} \vec{\mathbf{x}}}
\end{align}
where $N$ is the number of modes in a system, $\mathbf{x}$ is a vector of all of the quadrature moment operators such that $\mathbf{x} = \{ \vec{q}_{\alpha}, \vec{p}_{\alpha}, \vec{q}_{\beta}, \vec{p}_{\beta} \}$, and $C$ is some function of the quadrature moment operators. Using Wick's theorem, \ref{genint} becomes
\begin{align}
    \int d^{4N} \mathbf{x} C e^{-\frac{1}{2} \mathbf{x}^T \mathbf{A} \mathbf{x}} = \sqrt{\frac{(2 \pi)^{4N}}{\det{\mathbf{A}}}} \mathbf{W}(A,C,\vec{x}) \nonumber \\
    = \frac{(2\pi)^{2N}}{\sqrt{\det{\mathbf{A}}}} \mathbf{W}(A,C,\vec{x})
\end{align}
Where $\mathbf{W}(A,C,\vec{x})$ is the wick coupling of $C = (x_i x_j ... x_l)$ as defined for the real case (since we will operate in the phase-space basis), which is that 
\begin{align}
    &\mathbf{W}(A,C,\vec{x}) \nonumber \\
    &= \sum_{\text{all possible pairings P of } \{1,...,n\} } \langle x_{P_1} x_{P_2} \rangle ... \langle x_{P_{n-1}} x_{P_n} \rangle
\end{align}

\textit{The W function is available in genqo under the routine \texttt{W}, which ingests the A matrix and the symbolic vector of quadrature variables, which selects which of the terms in the Gaussian moment integral is being considered.}

\subsubsection{Alternative method: Calculation via the Hafnian}

The hafnian of an $n x n$ symetric matrix $A = A^T$ is defined as 
\begin{align}
    \text{haf}(A) = \sum_{M \in \text{PMP}(n)} \prod_{(i,j)\in M} A_{i,j}
\end{align}
Where PMP$(n)$ stands for the set of perfect matching permutations of n (even) objects. For example, for $n=4$, 
\begin{align}
    \text{PMP}(4) = \{(0,1)(2,3), (0,2)(1,3), (0,3)(1,2) \}
\end{align}
In our notation and in terms of Wick's theorem, the Hafnian calculation for $A$ is 
\begin{align}
    \text{haf}(A) = \mathbf{W}(A,\prod_{i=1}^n x_i, \vec{x})
\end{align}
In other words, it is the Wick coupling calculation over all of the elements of the basis vector. In the next section, we will provide a specific example for how this calculation is performed by taking sub-matrices of the case-specific $\mathbf{A}$ matrix. 

\onecolumngrid

\subsection{Probability of Generation}
\label{app:prob-fid}
\textit{To calculate the probability of generation as described in this section, use \texttt{calculate\_probability\_success}, which depends on the precomputed $\mathbf{A}_{pgen}$ matrix and W function}.

To verify that the approach presented in the preceding sections is correct, it is valuable to compare with the results of the cascaded source, as presented in \cite{dhara_heralded_2022}. The probability of generating the cascaded source state is 
\begin{align}
    P_{\rm gen} = \Tr(\Pi_e \rho_{z} )
\end{align}
Where 
\begin{align}
    \Pi_e &= \mathbf{I}^{\otimes 2} \otimes \outprod{n_3, n_4, n_5, n_6} \otimes \mathbf{I}^{\otimes 2}
\end{align}
Where $\{n_3, n_4, n_5. n_6\}$ is the heralding bell state click patterns, and $\rho_{z}$ (equation \ref{zstate}) is the state of the cascaded source in the coherent state basis. Therefore
\begin{align}
    \Tr(\Pi_e \rho_{zl}) &\propto \left( \prod_{i \in \{1,2,7,8 \} } \inprod{\beta_i \sqrt{\eta_i} }{\alpha_i \sqrt{\eta_i}} \right) \left( \prod_{j \in \{3,4,5,6 \} } \inprod{n_j}{\alpha_j \sqrt{\eta_j} } \inprod{\beta_j \sqrt{\eta_j} }{n_j} \right) \\
    \inprod{\beta_i \sqrt{\eta_i} }{\alpha_i \sqrt{\eta_i}} &= \exp \left( -\frac{\eta_i}{2} (|\alpha_i|^2 + |\beta_i|^2 - 2 \alpha_i \beta_i^*) \right) = \exp \left( {-\frac{1}{2}\left( |\alpha_j \sqrt{\eta_j}|^2 + |\beta_j \sqrt{\eta_j}|^2 \right)} \right) \exp \left( {\eta_i \alpha_i \beta_i^*} \right) \\
    \inprod{n_j}{\alpha_j \sqrt{\eta_j} } \inprod{\beta_j \sqrt{\eta_j} }{n_j} &= \exp \left( {-\frac{1}{2}\left( |\alpha_j \sqrt{\eta_j}|^2 + |\beta_j \sqrt{\eta_j}|^2 \right)} \right) \frac{(\alpha_j \beta_j^* \eta_j)^{n_j}}{n_j!}
\end{align}

\subsubsection{Calculating $\mathbf{A}_{pgen}$}
\textit{The math for calculating $\mathbf{A}_{pgen}$, is translated to code in Python under the functions \texttt{calculate\_k\_function\_matrix} and \texttt{calculate\_loss\_bsm\_matrix\_pgen}. It depends on the K-function matrix, as well as the loss parameters in each mode $\eta_i$.}

The matrix $\mathbf{A}_{pgen}$ is a modified form of the $\mathbf{A}$ matrix, whose modifications are specific to the probability of generation calculation as described in detail in this section. Recalling the $\mathbf{A}$ matrix for ZALM (\ref{amat}) is 
\begin{align}
    &\exp \left[ -\frac{1}{2} \vec{x} A \vec{x}^T \right] = \exp\left[-\frac{1}{2}\vec{x}(\mathcal{B}_{\alpha} \oplus \mathcal{B}^*_{\beta})\vec{x}^T \right] \\ \nonumber
    &\times \exp \left[ -\frac{1}{2}\left( \sum_{i = 1}^{8} (\eta_i - 1)(q_{\alpha_i} + i p_{\alpha_i})(q_{\beta_i} - i p_{\beta_i}) \right) \right] \\ \nonumber 
    &\times \exp \left[ -\frac{1}{2}\left( \sum_{i = 1}^{8} \frac{1}{2}(|q_{\alpha_i}|^2 +  |p_{\alpha_i}|^2 + |q_{\beta_i}|^2 +  |p_{\beta_i}|^2) \right) \right] 
\end{align}
For calculating the probability of generation, the $\mathbf{A}$ matrix will be a modified such that
\begin{align}
    \exp \left[ -\frac{1}{2} \vec{x} \mathbf{A}_{pgen} \vec{x}^T \right] &= \exp \left[ -\frac{1}{2} \vec{x} \mathbf{A} \vec{x}^T \right] \exp \left[ -\frac{1}{2} {\sum_{i \in \{1,2,7,8 \} }  -2\eta_i \alpha_i \beta_i^*} \right]  \\
    &= \exp\left[-\frac{1}{2}\vec{x}(\mathcal{B}_{\alpha} \oplus \mathcal{B}^*_{\beta})\vec{x}^T \right] \exp \left[ -\frac{1}{2}\left( \sum_{i = 1}^{8} \frac{1}{2} (|q_{\alpha_i}|^2 +  |p_{\alpha_i}|^2 + |q_{\beta_i}|^2 +  |p_{\beta_i}|^2) \right) \right] \nonumber \\  
    &\times \exp \left[ -\frac{1}{2}\left( \sum_{i = 3}^{6}  (\eta_i - 1)(q_{\alpha_i} + i p_{\alpha_i})(q_{\beta_i} - i p_{\beta_i}) \right) \right] \nonumber \\ 
    &\times \exp \left[ -\frac{1}{2}\left( \sum_{i \in \{1,2,7,8\} } \left(- 1 \right)(q_{\alpha_i} + i p_{\alpha_i})(q_{\beta_i} - i p_{\beta_i}) \right) \right] 
\end{align}

\textbf{Important Notice:} the modes where the detection is composed of the identity matrix will incorporate cross terms in the $\mathbf{A}$ matrix which are not found in a standard detection scheme. These cross terms are significant and must be incorporated correctly. Consider how the $\mathbf{A}$ matrix terms are written in summation notation for just the single mode case
\begin{align}
    \begin{bmatrix}
        q_1 & p_1
    \end{bmatrix} \begin{bmatrix}
        A_{11} & A_{12} \\
        A_{21} & A_{22}
    \end{bmatrix} \begin{bmatrix}
        q_1 \\
        p_1
    \end{bmatrix} = A_{11} q_1^2 + A_{22} p_1^2 + 2 A_{12} q_1 p_1
\end{align}
If we take the cross terms as we have calculated them, they will be off by a factor of $\frac{1}{2}$. So, when incorporating these cross terms into the $\mathbf{A}_{pgen}$ matrix in Python, it is essential to include this extra term. 

Therefore, using Wick's Theorem, this can be simplified as
\begin{align} \label{pgen}
    \Tr(\Pi_e \rho_{zl} ) &= \frac{\eta_b^2}{(2\pi)^{16} (\det{\Gamma})^{1/4} (\det{\Gamma^*})^{1/4}} \int d^{16} \vec{x}^{(l)}_{\alpha}  d^{16} \vec{x}^{(l)}_{\beta} (\alpha_3 \beta_3^* \alpha_4 \beta_4^*)   \exp \left[ -\frac{1}{2} \vec{x} \mathbf{A}_{pgen} \vec{x}^T \right] \\
    &= \frac{\eta_b^2 (2\pi)^{16}}{(2\pi)^{16} (\det{\Gamma})^{1/4} (\det{\Gamma^*})^{1/4} \sqrt{\det{\mathbf{A}_{pgen}}} } \mathbf{W}(\mathbf{A}_{pgen}, \alpha_3 \beta_3^* \alpha_4 \beta_4^*) \\
    &= \frac{\eta_b^2}{ (\det{\Gamma})^{1/4} (\det{\Gamma^*})^{1/4} \sqrt{\det{\mathbf{A}_{pgen}}} } \mathbf{W}(\mathbf{A}_{pgen}, \alpha_3 \beta_3^* \alpha_4 \beta_4^*)
\end{align}
Where we describe the details of $\mathbf{W}(...)$ in section \ref{app:wick}

We know that calculating the probability of generation is a matter of calculating  $\mathbf{W}(A_{pgen}, \alpha_3 \beta_3^* \alpha_4 \beta_4^*, \vec{x} )$. This will have sub elements after Wick coupling that look like
\begin{align}
    \mathbf{W}(A_{pgen}, p_{\alpha,3} q_{\beta,3} p_{\alpha,4} q_{\beta,4}, \vec{x} )
\end{align}
If we create some submatrix $A_{sub}$ which is derived from $A_{pgen}$ by only including the elements that correspond to $\{ p_{\alpha,3}, q_{\beta,3}, p_{\alpha,4}, q_{\beta,4}\}$, such that $A_{sub}$ has the basis vector $\vec{x}_{sub} = \{p_{\alpha,3}, q_{\beta,3}, p_{\alpha,4}, q_{\beta,4} \}$, then 
\begin{align}
    \mathbf{W}(A_{pgen}, p_{\alpha,3} q_{\beta,3} p_{\alpha,4} q_{\beta,4}, \vec{x} ) = \text{haf}(A_{sub})
\end{align}

\subsubsection{Dark Counts}
Considering dark counts will introduce a slight modification to the C coefficient while keeping the A matrix unchanged. For calculating the probability of generation, the formula for dark counts is 
\begin{align}
    P_{\rm gen}(P_d) &= \frac{1}{ (\det{\Gamma})^{1/4} (\det{\Gamma^*})^{1/4} } \nonumber \\
    &[ \frac{1}{\sqrt{\det{\mathbf{A}_{pgen}} }} \eta_b^2 (1 - P_d)^2 \mathbf{W}(\mathbf{A}_{pgen}, \alpha_3 \beta_3^* \alpha_4 \beta_4^*) \nonumber \\ 
     &+ \frac{1}{\sqrt{\det{\mathbf{A}_{pgen,1}} }}\eta_b P_d (1 - P_d) \mathbf{W}(\mathbf{A}_{pgen,1}, \alpha_3 \beta_3^*) \nonumber \\
     &+ \frac{1}{\sqrt{\det{\mathbf{A}_{pgen,2}} }}\eta_b P_d (1 - P_d) \mathbf{W}(\mathbf{A}_{pgen,2}, \alpha_4 \beta_4^*)  \nonumber \\ 
     &+  \frac{1}{\sqrt{\det{\mathbf{A}_{pgen,3}} }} P_d^2 \mathbf{W}(\mathbf{A}_{pgen,3}, 1) ]
\end{align}
Notice that as different click patterns are accepted as correct due to dark clicks it is necessary to change the A matrix just slightly, which is why each term is dependent upon different A matrices. 

\subsubsection{Fidelity Calculation}
\textit{The math for calculating the fidelity is translated to code in Python under the functions \texttt{calculate\_fidelity}. It depends on the $\mathbf{A}$ matrix, the $\mathbf{A}_{pgen}$ matrix, the W function, and the loss parameters $\eta_t$, $\eta_d$, $\eta_b$ }

The fidelity of the state with respect to the bell state is
\begin{align}
    F &= \frac{\bra{\xi} \rho_{z} \ket{\xi} }{\Tr(\rho_{z})} \\
    \ket{\xi} &= \frac{1}{\sqrt{2}} \left( \ket{1,0,n_3,n_4,n_5,n_6,0,1} - \ket{0,1,n_3,n_4,n_5,n_6,1,0} \right)
\end{align}
Functionally, this can be calculated in similar fashion to the probability of generation, where the projection operator will change slightly.
\begin{align}
    \Tr(\rho_{z,b}) = \Tr(\Pi_e \rho_{z})
\end{align}
Where $\rho_{z,b}$ is $\rho_{z}$ following the measurement on the bell state modes. Notice that this is the probability of generation. Therefore, the fidelity simplifies to
\begin{align}
    F &= \frac{1}{2}\frac{\bra{1,0,0,1} \rho_{z,b} \ket{1,0,0,1} + \bra{1,0,0,1} \rho_{z,b} \ket{0,1,1,0} + \bra{0,1,1,0} \rho_{z} \ket{1,0,0,1} + \bra{0,1,1,0} \rho_{z,b} \ket{0,1,1,0} }{ \Tr(\Pi_e \rho{z}) } \nonumber \\
    &= \left( \frac{1}{2} \frac{(\eta_d \eta_t \eta_b)^2}{\sqrt{\det{\mathbf{A}}}} \frac{\sqrt{\det{\mathbf{A}_{pgen}}}}{\eta_b^2} \frac{1}{ \mathbf{W}(\mathbf{A}_{pgen}, \alpha_3 \beta_3^* \alpha_4 \beta_4^*) } \right)  \times \nonumber \\ 
    &\left( \mathbf{W}(\mathbf{A}, \alpha_1 \alpha_3 \alpha_4 \alpha_8 \beta_1^* \beta_3^* \beta_4^* \beta_8^*) + \mathbf{W}(\mathbf{A}, \alpha_1 \alpha_3 \alpha_4 \alpha_8 \beta_2^* \beta_3^* \beta_4^* \beta_7^*) + \mathbf{W}(\mathbf{A}, \alpha_2 \alpha_3 \alpha_4 \alpha_7 \beta_1^* \beta_3^* \beta_4^* \beta_8^*) + \mathbf{W}(\mathbf{A}, \alpha_2 \alpha_3 \alpha_4 \alpha_7 \beta_2^* \beta_3^* \beta_4^* \beta_7^*) \right) \nonumber \\
    &= \left( \frac{(\eta_d \eta_t)^2 \sqrt{\det{\mathbf{A}_{pgen}}} }{2 \sqrt{\det{\mathbf{A}}}} \frac{1}{ \mathbf{W}(\mathbf{A}_{pgen}, \alpha_3 \beta_3^* \alpha_4 \beta_4^*) } \right) \times \nonumber \\ 
    &\left( \mathbf{W}(\mathbf{A}, \alpha_1 \alpha_3 \alpha_4 \alpha_8 \beta_1^* \beta_3^* \beta_4^* \beta_8^*) + \mathbf{W}(\mathbf{A}, \alpha_1 \alpha_3 \alpha_4 \alpha_8 \beta_2^* \beta_3^* \beta_4^* \beta_7^*) + \mathbf{W}(\mathbf{A}, \alpha_2 \alpha_3 \alpha_4 \alpha_7 \beta_1^* \beta_3^* \beta_4^* \beta_8^*) + \mathbf{W}(\mathbf{A}, \alpha_2 \alpha_3 \alpha_4 \alpha_7 \beta_2^* \beta_3^* \beta_4^* \beta_7^*) \right)
\end{align}

The analysis so far has only considered the photon-photon state produced by a cascaded source. The next step is to consider the state of two spin quantum memories, after this state has been distributed and loaded. The next section will describe the process by which we model this loading, and how the final spin-spin state can be calculated using the same tools as have already been described. 

\onecolumngrid

\subsection{Incorporating Duan-Kimble style Quantum Memories}
\label{app:memory}
Essential to this modeling is the considerations of quantum memories that store the arriving entangled bell pairs. The quantum memories are atomic qubits where the quantum state might be the spin state of an electron in a color center, but the Duan-Kimble protocol~\cite{duan_scalable_2004} is agnostic to the type of qubit and rather focuses on the fact that a cavity QED system enables light-matter coupling to realize a control-Z gate. In summary, loading the photon-photon entangled state into the quantum memory spin state via the Duan-Kimble protocol is achieved by
\begin{enumerate}
    \item The photonic state interacts with the spin state via a cavity interaction, resulting in a control-Z gate between the state of the spin and the state of the photon.
    \item The photonic state enters a polarizing beamsplitter, where each pair of modes interferes
    \item The photonic state is detected with photon detectors
\end{enumerate}
For the purpose of our modeling we will treat these qubit basis states as bosonic modes. The interaction will be treated as a cross-Kerr unitary between the photonic source state and simulated atomic qubit, treated as a dual-rail bosonic mode.

\subsubsection{Quantum Memory Interactions} \label{qmems}
We begin by describing the non-gaussian controlled-z operation between the bosonic modes generated by the entanglement distribution sources and the quantum memory simulated-bosonic modes. Recall that in the Schrodinger picture of quantum mechanics, a general unitary interaction can be applied to a quantum state by 
\begin{align}
    \hat{U}(t) \ket{\Psi} = \exp(i \hat{H} t) \ket{\Psi}
\end{align}
Where $\hat{H}$ is the Hamiltonian that represents the interaction. Recall that the series representation of an exponential is
\begin{align}
    \exp(x) = \sum_{n=0}^\infty \frac{x^n}{n!} = 1 + x + \frac{x^2}{2} + ...
\end{align}
The series representation of the exponential can be used to apply the unitary operation to the quantum state. For example, recalling the fundamental operator relationships \ref{operators}, the phase operator is defined as
\begin{align}
    U_{\theta} = e^{i \theta \hat{N}} = e^{i \theta \hat{a}^{\dagger} \hat{a}}
\end{align}
When applied to the 0 and 1 fock states, this results in 
\begin{align}
    U_{\theta} \ket{0} &= \ket{0} \\
    U_{\theta} \ket{1} &= e^{i \theta} \ket{1} 
\end{align}
When applied to a coherent state $\ket{\alpha}$
\begin{align}
    \hat{U}_{\theta} \ket{\alpha} &= e^{i \theta \hat{a}^{\dagger} \hat{a}} \ket{\alpha} \nonumber \\ 
    &= (1 + (i \theta \hat{a}^{\dagger} \hat{a}) + \frac{(i \theta \hat{a}^{\dagger} \hat{a})^2}{2} + ...) \ket{\alpha} \nonumber \\
    &= (1 + (i \theta \hat{a}^{\dagger} \hat{a}) + \frac{(i \theta \hat{a}^{\dagger} \hat{a})^2}{2} + ...) \sum_{n=0}^{\infty} (\frac{e^{\frac{-|\alpha|^2}{2}} \alpha^n}{\sqrt{n!}}) \ket{n} \nonumber \\
    &= \sum_{n=0}^{\infty} (\frac{e^{\frac{-|\alpha|^2}{2}} \alpha^n}{\sqrt{n!}}) (1 + (i \theta \hat{a}^{\dagger} \hat{a}) + \frac{(i \theta \hat{a}^{\dagger} \hat{a})^2}{2} + ...)\ket{n} \nonumber \\
    &= \sum_{n=0}^{\infty} (\frac{e^{\frac{-|\alpha|^2}{2}} \alpha^n}{\sqrt{n!}}) (1 + (i \theta n) + \frac{(i \theta n)^2}{2} + ...)\ket{n} \nonumber \\
    &= \sum_{n=0}^{\infty} (\frac{e^{\frac{-|\alpha|^2}{2}} \alpha^n}{\sqrt{n!}}) e^{i \theta n} \ket{n} \nonumber \\
    &= \sum_{n=0}^{\infty} (\frac{e^{\frac{-|\alpha|^2}{2}} (\alpha e^{i \theta})^n}{\sqrt{n!}}) \ket{n} \nonumber \\
    &= \ket{\alpha e^{i \theta}}
\end{align}
A control-z operation between two modes a and b can be performed using a cross-Kerr unitary \cite{imoto_quantum_1985} where $\theta = \pi$ of the form
\begin{align}
    U_{ck} = e^{i \pi \hat{a}^{\dagger} \hat{a} \hat{b}^{\dagger} \hat{b}}
\end{align}
As applied between a single-rail qubit and a coherent state, the result is
\begin{align}
    U_{ck}(\frac{\ket{0} + \ket{1}}{\sqrt{2}})\ket{\alpha} = \frac{1}{\sqrt{2}}(\ket{0} \ket{\alpha} + \ket{1} \ket{-\alpha})
\end{align}
This relation can be used to interact the source state, which using the K-function formalism is represented as an integral over coherent states, with the simulated atomic memories, which are represented as dual-rail qubits. First, define
\begin{align}
    \rho_M &= \ket{\phi} \bra{\phi} \\
    \ket{\phi} &= \frac{1}{\sqrt{2}}(\ket{1;0} + \ket{1;0})
\end{align}
Therefore, the state of our system prior to interaction is
\begin{align}
    \rho = \rho_{zl} \otimes \rho_M \otimes \rho_M
\end{align}
Applying cross-Kerr unitaries as interactions of the source mode with the memory modes
\begin{align}
    \rho_{z,int} =  U_{ck_{8,4}} U_{ck_{2,2}} \rho U^{\dagger}_{ck_{2,2}} U^{\dagger}_{ck_{8,4}}
\end{align}
Where
\begin{align}
    U_{ck_{i,j}} = e^{i \pi \hat{a}_i^{\dagger} \hat{a}_i \hat{b}_j^{\dagger} \hat{b}_j}
\end{align}
With $\hat{a}_i^{\dagger}$ and $\hat{b}_i^{\dagger}$ being the creation operators of the $i$th source mode and memory mode respectively. Therefore
\begin{align}
    \rho_{z,int} = \int d^{16} \vec{x_{\alpha}} d^{16} \vec{x_{\beta}} K(\vec{\alpha}) K(\vec{\beta})^* \mathcal{G}(\vec{\alpha}, \vec{\beta}, \vec{\eta}) \ket{\vec{\alpha}_{E}} \bra{\vec{\beta}_{E}} 
\end{align}
Where
\begin{align}
    \ket{\vec{\alpha}_{E}} &= \frac{1}{2} U_{ck_{4,8}} U_{ck_{2,2}} \ket{\vec{\alpha}_L}(\ket{0,1} + \ket{1,0})(\ket{0,1} + \ket{1,0}) \nonumber \\
    &= \frac{1}{2}( \ket{0,1} \ket{0,1} \bigotimes_{i = 1}^8 \begin{cases}
        \ket{-\alpha_i \sqrt{\eta_i}} \text{ for } i = 2,8 \\
        \ket{\alpha_i \sqrt{\eta_i}} \text{ for } i = 1,7 \\
        \ket{\alpha_i \sqrt{\eta_i}} \text{ for } i = 3,4,5,6
    \end{cases} \nonumber \\
    &+ \ket{0,1} \ket{1,0} \bigotimes_{i = 1}^8\begin{cases}
        \ket{-\alpha_i \sqrt{\eta_i}} \text{ for } i = 2 \\
        \ket{\alpha_i \sqrt{\eta_i}} \text{ for } i = 1,7,8 \\
        \ket{\alpha_i \sqrt{\eta_i}} \text{ for } i = 3,4,5,6
    \end{cases} \nonumber \\
    &+ \ket{1,0} \ket{0,1} \bigotimes_{i = 1}^8\begin{cases}
        \ket{-\alpha_i \sqrt{\eta_i}} \text{ for } i = 8 \\
        \ket{\alpha_i \sqrt{\eta_i}} \text{ for } i = 1,2,7 \\
        \ket{\alpha_i \sqrt{\eta_i}} \text{ for } i = 3,4,5,6
    \end{cases}\nonumber \\
    &+ \ket{1,0} \ket{1,0} \bigotimes_{i = 1}^8\begin{cases}
        \ket{\alpha_i \sqrt{\eta_i}} \text{ for } i = 1,2,7,8 \\
        \ket{\alpha_i \sqrt{\eta_i}} \text{ for } i = 3,4,5,6
    \end{cases})
\end{align}
Next, each pair of modes will pass through a polarizing beam splitter.

\subsubsection{Interference to complete Encoding} \label{beamsp}
Consider two coherent states $\ket{\alpha}$ and $\ket{\beta}$
\begin{align}
    &\ket{\alpha} \ket{\beta} \nonumber \\
    &= (e^{(-|\alpha|^2)/2}) \sum_{n = 0}^{\infty} \frac{(\alpha)^n}{\sqrt{n!}} \ket{n}) (e^{(-|\beta|^2)/2}) \sum_{m = 0}^{\infty} \frac{(\beta)^m}{\sqrt{m!}} \ket{m}) \nonumber \\
    &= (e^{(-|\alpha|^2)/2}) \sum_{n = 0}^{\infty} \frac{(\alpha)^n}{\sqrt{n!}} (\hat{a}^n) \ket{0}) (e^{(-|\beta|^2)/2}) \sum_{m = 0}^{\infty} \frac{(\beta)^m}{\sqrt{m!}} (\hat{b}^m) \ket{0}) \nonumber \\
    &= e^{(-|\alpha|^2)/2} e^{(-|\beta|^2)/2} \sum_{n = 0}^{\infty} \sum_{m = 0}^{\infty} \frac{(\alpha)^n}{\sqrt{n!}} \frac{(\beta)^m}{\sqrt{m!}} (\hat{a}^n) (\hat{b}^m) \ket{0} \ket{0}
\end{align}
We know that the polarizing beam splitter will transform the creation operators such that
\begin{align}
    \hat{a} &\rightarrow \frac{\hat{a} + \hat{b}}{\sqrt{2}} \\
    \hat{b} &\rightarrow \frac{\hat{a} - \hat{b}}{\sqrt{2}}
\end{align}
Therefore
\begin{align}
    U_{pbs} \ket{\alpha} \ket{\beta} &= e^{(-|\alpha|^2)/2} e^{(-|\beta|^2)/2} \sum_{n = 0}^{\infty} \sum_{m = 0}^{\infty} \frac{(\alpha)^n}{\sqrt{n!}} \frac{(\beta)^m}{\sqrt{m!}} (\frac{\hat{a} + \hat{b}}{\sqrt{2}})^n (\frac{\hat{a} - \hat{b}}{\sqrt{2}})^m \ket{0} \ket{0} \\
    &= e^{(-|\alpha|^2)/2} e^{(-|\beta|^2)/2} \sum_{n = 0}^{\infty} \sum_{m = 0}^{\infty} \frac{(\alpha)^n}{\sqrt{n!}} \frac{(\beta)^m}{\sqrt{m!}} \frac{1}{(\sqrt{2})^{n+m}} (\hat{a} + \hat{b})^n (\hat{a} - \hat{b})^m \ket{0} \ket{0} \\ \nonumber 
\end{align}
Consider the binomial theorem
\begin{align}
    (x+y)^n = \sum_{k = 0}^{n} \binom{n}{k} x^{n-k} y^k = \sum_{k = 0}^{n} \binom{n}{k} x^{k} y^{n-k} 
\end{align}
So, 
\begin{align}
    (\hat{a} + \hat{b})^{n} &= \sum_{k = 0}^n \binom{n}{k} \hat{a}^{n-k} \hat{b}^{k} \\
    (\hat{a} - \hat{b})^{m} &= \sum_{l = 0}^m \binom{m}{l} \hat{a}^{m-l} (-1)^{l} \hat{b}^{l}
\end{align}
Therefore, our expression becomes
\begin{align}
    U_{pbs} \ket{\alpha} \ket{\beta} &= e^{(-|\alpha|^2)/2} e^{(-|\beta|^2)/2} \sum_{n = 0}^{\infty} \sum_{m = 0}^{\infty} \frac{(\alpha)^n}{\sqrt{n!}} \frac{(\beta)^m}{\sqrt{m!}} \frac{1}{(\sqrt{2})^{n+m}} \sum_{k = 0}^n \sum_{l = 0}^m \binom{n}{k} \binom{m}{l} (-1)^{l} \hat{a}^{n+m-k-l} \hat{b}^{k+l} \ket{0} \ket{0} \nonumber \\
    &= e^{(-|\alpha|^2)/2} e^{(-|\beta|^2)/2} \sum_{n = 0}^{\infty} \sum_{m = 0}^{\infty} \frac{(\alpha)^n}{\sqrt{n!}} \frac{(\beta)^m}{\sqrt{m!}} \frac{1}{(\sqrt{2})^{n+m}} \sum_{k = 0}^n \sum_{l = 0}^m \binom{n}{k} \binom{m}{l} (-1)^{l} \ket{n+m-k-l} \ket{k+l} \\
\end{align}
This expression will simplify as we consider photon detection on these coherent state modes. 

\subsubsection{Detection} \label{det}
For modeling detection, it is desireable to include the non-idealities of loss and detector dark clicks. Detector loss can be incorporated using the same prescription as transmission loss (as described in \ref{loss}), just replace $\eta$ with $\eta_d$. Detector dark clicks can be incorporated by calculating the probability of incorrect click patterns that would be accepted in the presence of detector dark clicks. 

For the Bell state measurement modes, the desired click patterns are known exactly from \citep{dhara_heralded_2022}. As for the remaining modes, heralded photons will collapse the state of the memories, and the click patterns for these detectors are known based on the encoding technique. For example, with Duan-Kimble memory encoding we know that acceptable click patterns are those where only one click is detected for each pair of source modes. For further details, see the examples that follow. 

In many cases, the detection situation that will be considered are
\begin{itemize}
    \item $\bra{1} \bra{0} U_{pbs} \ket{\alpha} \ket{\beta}$
    \item $\bra{0} \bra{1} U_{pbs} \ket{\alpha} \ket{\beta}$
    \item $\bra{0} \bra{0} U_{pbs} \ket{\alpha} \ket{\beta}$
\end{itemize}
First, consider the most simple case
\begin{align}
    \bra{0} \bra{0} U_{pbs} \ket{\alpha} \ket{\beta} &= e^{(-|\alpha|^2)/2} e^{(-|\beta|^2)/2}
\end{align}
Next, for the single-photon detection patterns, we consider the overlap in products, which result in
\begin{align}
    n+m-l-k &= 1 \\
    l+k &= 0 \rightarrow l = -k
\end{align}
Because $k$ cannot be negative, $l = k = 0$. So, therefore $n+m = 1$, and so either $n=0, m=1$ or $n=1, m=0$. So, 
\begin{align}
    &\bra{1} \bra{0} U_{pbs} \ket{\alpha} \ket{\beta} \nonumber \\
    &= \frac{1}{\sqrt{2}} e^{(-|\alpha|^2)/2} e^{(-|\beta|^2)/2} \left( \binom{1}{0} \binom{0}{0} \alpha + \binom{0}{0} \binom{1}{0} \beta \right) \nonumber \\
    &= e^{(-|\alpha|^2)/2} e^{(-|\beta|^2)/2} \left( \frac{\alpha + \beta}{\sqrt{2}} \right)
\end{align}
Now consider $\bra{0} \bra{1} U_{pbs} \ket{\alpha} \ket{\beta}$. Either $l = 1, k = 0$ or $l = 0, k = 1$, and similarly $n+m-1=0$ so $n+m = 1$, therefore either $n = 1, m=0$ or $n=0,m=1$ and there are 4 terms that result
\begin{enumerate}
    \item $n=0, m=1, l=1, k=0$
    \item $n=0, m=1, l=0, k=1$
    \item $n=1, m=0, l=1, k=0$
    \item $n=1, m=0, l=0, k=1$
\end{enumerate}
Resulting in 
\begin{align}
    \bra{0} \bra{1} U_{pbs} \ket{\alpha} \ket{\beta}
    &= e^{(-|\alpha|^2)/2} e^{(-|\beta|^2)/2} \left( \frac{\alpha - \beta}{\sqrt{2}} \right)
\end{align}

\subsubsection{Calculating the Spin-Spin Density Matrix}
The spin-spin density matrix for the cascaded source is therefore
\begin{align}
    \rho_{A,B} &= U_{pbs,(7,8)} U_{pbs,(1,2)} \rho_{z,int} U_{pbs,(1,2)}^{\dagger} U_{pbs,(7,8)}^{\dagger} 
\end{align}
Which simplifies as
\begin{align}
    \rho_{A,B} &= \left(\frac{1}{2}\right) \left(\frac{1}{2}\right) \int d^{16} \vec{x_{\alpha}} d^{16} \vec{x_{\beta}} K(\vec{\alpha}) K(\vec{\beta})^* \mathcal{G}(\vec{\alpha}, \vec{\beta}, \vec{\eta}) \exp\left((-\frac{1}{2}\sum_{i=1}^8 (|\alpha_i \sqrt{\eta_i}|^2 + |\beta_i \sqrt{\eta_i}|^2))\right) \left( \prod_{i = 3}^6 \frac{(\alpha_i \beta_i^* \eta_i)^{n_i}}{n_i!} \right) \vec{A} \vec{B} \\
    \vec{A} = &\ket{0,1} \ket{0,1} \left(\frac{\alpha_1 \sqrt{\eta_1} - \alpha_2\sqrt{\eta_2}}{\sqrt{2}} \right)^{n_1} \left(\frac{\alpha_1 \sqrt{\eta_1} + \alpha_2\sqrt{\eta_2}}{\sqrt{2}} \right)^{n_2} \left(\frac{\alpha_7 \sqrt{\eta_7} - \alpha_8\sqrt{\eta_8}}{\sqrt{2}} \right)^{n_7} \left(\frac{\alpha_7 \sqrt{\eta_7} + \alpha_8\sqrt{\eta_8}}{\sqrt{2}} \right)^{n_8} \nonumber \\
    + &\ket{0,1} \ket{1,0} \left(\frac{\alpha_1 \sqrt{\eta_1} - \alpha_2\sqrt{\eta_2}}{\sqrt{2}} \right)^{n_1} \left(\frac{\alpha_1 \sqrt{\eta_1} + \alpha_2\sqrt{\eta_2}}{\sqrt{2}} \right)^{n_2} \left(\frac{\alpha_7 \sqrt{\eta_7} + \alpha_8\sqrt{\eta_8}}{\sqrt{2}} \right)^{n_7} \left(\frac{\alpha_7 \sqrt{\eta_7} - \alpha_8\sqrt{\eta_8}}{\sqrt{2}} \right)^{n_8} \nonumber \\
    + &\ket{1,0} \ket{0,1} \left(\frac{\alpha_1 \sqrt{\eta_1} + \alpha_2\sqrt{\eta_2}}{\sqrt{2}} \right)^{n_1} \left(\frac{\alpha_1 \sqrt{\eta_1} - \alpha_2\sqrt{\eta_2}}{\sqrt{2}} \right)^{n_2} \left(\frac{\alpha_7 \sqrt{\eta_7} - \alpha_8\sqrt{\eta_8}}{\sqrt{2}} \right)^{n_7} \left(\frac{\alpha_7 \sqrt{\eta_7} + \alpha_8\sqrt{\eta_8}}{\sqrt{2}} \right)^{n_8} \nonumber \\
    + &\ket{1,0} \ket{1,0} \left(\frac{\alpha_1 \sqrt{\eta_1} + \alpha_2\sqrt{\eta_2}}{\sqrt{2}} \right)^{n_1} \left(\frac{\alpha_1 \sqrt{\eta_1} - \alpha_2\sqrt{\eta_2}}{\sqrt{2}} \right)^{n_2} \left(\frac{\alpha_7 \sqrt{\eta_7} + \alpha_8\sqrt{\eta_8}}{\sqrt{2}} \right)^{n_7} \left(\frac{\alpha_7 \sqrt{\eta_7} - \alpha_8\sqrt{\eta_8}}{\sqrt{2}} \right)^{n_8} \\
\end{align}
Where $\vec{B}$ is similar to $\vec{A}$ except all $\alpha_i$'s become $\beta_i$'s and the memory basis kets become basis bras. Also, $\ket{n_1,n_2,n_3,n_4,n_5,n_6,n_7,n_8}$ is the measurement performed on the photonic modes, assuming that for $i \in \{1,2,7,8\}$ $n_i \leq 1$. 

We know that there are many detection patterns which will herald the two events necessary for memory loading: 1) bell state heralding at the source; 2) memory load success heralding. For simulation purposes, we will consider the pattern $\ket{1,0,1,1,0,0,1,0}$

Notice that the $A$ matrix, as defined for the general detection upon the source photonic state, does not change following quantum memory loading. All that changes is the $C$ coefficient. At this point we are ready to consider this state in the presence of dark clicks. 

\subsubsection{Incorporating Dark Clicks}

To herald the generation of a spin-spin bell state in the presence of dark clicks, the state which we will calculate is
\begin{align}
    \rho_{A,B}(P_d) &= (1 - P_d)^8 \rho_{A,B} + P_d(1-P_d)^7 \left( \sum_{i = 1}^4\sigma_{1,i} \right) + P_d^2(1-P_d)^6 \left( \sum_{i = 1}^6\sigma_{2,i} \right) + P_d^3(1-P_d)^5 \left( \sum_{i = 1}^4\sigma_{3,i} \right) + P_d^4(1-P_d)^4 \sigma_{4,1} 
\end{align}
Where $\vec{n} = \{n_1,n_2,n_3,n_4,n_5,n_6,n_7,n_8\}$
\begin{itemize}
    \item $\sigma_{1,1}$ occurs when $\vec{n} = \{ 0,0,1,1,0,0,1,0 \}$
    \item $\sigma_{1,2}$ occurs when $\vec{n} = \{ 1,0,0,1,0,0,1,0 \}$
    \item $\sigma_{1,3}$ occurs when $\vec{n} = \{ 1,0,1,0,0,0,1,0 \}$
    \item $\sigma_{1,4}$ occurs when $\vec{n} = \{ 1,0,1,1,0,0,0,0 \}$
    \item $\sigma_{2,1}$ occurs when $\vec{n} = \{ 0,0,0,1,0,0,1,0 \}$
    \item $\sigma_{2,2}$ occurs when $\vec{n} = \{ 0,0,1,0,0,0,1,0 \}$
    \item $\sigma_{2,3}$ occurs when $\vec{n} = \{ 0,0,1,1,0,0,0,0 \}$
    \item $\sigma_{2,4}$ occurs when $\vec{n} = \{ 1,0,0,0,0,0,1,0 \}$
    \item $\sigma_{2,5}$ occurs when $\vec{n} = \{ 1,0,0,1,0,0,0,0 \}$
    \item $\sigma_{2,6}$ occurs when $\vec{n} = \{ 1,0,1,0,0,0,0,0 \}$
    \item $\sigma_{3,1}$ occurs when $\vec{n} = \{ 1,0,0,0,0,0,0,0 \}$
    \item $\sigma_{3,2}$ occurs when $\vec{n} = \{ 0,0,1,0,0,0,0,0 \}$
    \item $\sigma_{3,3}$ occurs when $\vec{n} = \{ 0,0,0,1,0,0,0,0 \}$
    \item $\sigma_{3,4}$ occurs when $\vec{n} = \{ 0,0,0,0,0,0,1,0 \}$
    \item $\sigma_{4,1}$ occurs when $\vec{n} = \{ 0,0,0,0,0,0,0,0 \}$
\end{itemize}


\end{document}